\documentclass[preprint,authoryear,12pt]{elsarticle}

\usepackage{natbib}

\usepackage{graphics}
\usepackage{graphicx}
\usepackage{epsfig}
\usepackage{color}           

\usepackage{amssymb}

\renewcommand{\vec}[1]{ {\mathbf #1} }

\newcommand{\xx}{ \vec x}

\newcommand{\aap}{{\it Astron. Astrophys. }}

\newcommand{\apj}{{\it Astrophys. J. }}
\newcommand{\apjl}{{\it Astrophys. J. Lett. }}

\newcommand{\grl}{{\it Geophys. Res. Lett. }}

\newcommand{\jgr}{{\it J. Geophys. Res. }}

\newcommand{\solphys}{{\it Solar Phys. }}

\newcommand{\ssr}{{\it Space Sci. Rev. }}

\newcommand{\na}{{\it New Astronomy }}
\begin{document}
\begin{frontmatter}
\title{A Multiwavelength Study of Eruptive Events on January 23, 2012 Associated with a Major Solar Energetic Particle Event}
\author{N. C. Joshi\corref{cor}\fnref{footnote1}}
\address{Aryabhatta Research Institute of Observational Sciences (ARIES), Manora Peak, Nainital-263 129, India}
\cortext[cor]{Corresponding author}
\fntext[footnote1]{Ph: +91-5942-233735; Fax: +91-5942-233439}
\ead{navin@aries.res.in; njoshi98@gmail.com}
\author{W. Uddin\fnref{footnote1}}
\address{Aryabhatta Research Institute of Observational Sciences (ARIES), Manora Peak, Nainital-263 129, India}
\ead{wahab@aries.res.in}
\author{A.K. Srivastava\fnref{footnote1}}
\address{Aryabhatta Research Institute of Observational Sciences (ARIES), Manora Peak, Nainital-263 129, India}
\ead{aks@aries.res.in}
\author{R. Chandra\fnref{footnote2}}
\address{Department of Physics, DSB Campus, Kumaun University, Nainital-263 002, India}
\fntext[footnote2]{Ph: +91-5942-237450, Fax: +91-5942-237450}
\ead{rchandra.ntl@gmail.com}
\author{N. Gopalswamy\fnref{footnote3}}
\address{NASA Goddard Space Flight Center, Greenbelt, MD 20771, USA}
\fntext[footnote3]{Ph: +1-301-286-5885, Fax: +1-301-286-7194}
\ead{Nat.Gopalswamy@nasa.gov}
\author{P.K. Manoharan\fnref{footnote4}}
\address{Radio Astronomy Centre, Tata Institute of Fundamental Research, Ooty-643 001, India}
\fntext[footnote4]{Ph: +91-423-2550334, Fax: +91-423-2550135}
\ead{mano@ncra.tifr.res.in}
\author{M.J. Aschwanden\fnref{footnote5}}
\address{Lockheed Martin Solar and Astrophysics Laboratory, Palo Alto, CA 94304, USA}
\fntext[footnote5]{Ph: +1-650-424-4001, Fax: +1-650-424-3994}
\ead{aschwanden@lmsal.com}
\author{D.P. Choudhary\fnref{footnote6}}
\address{Deptartment of Physics and Astrophysics, California State University, Northridge, CA 91330-8268, USA}
\fntext[footnote6]{Ph: +1-818-677-7113; Fax: +1-818-677-3234}
\ead{debiprasad.choudhary@csun.edu}
\author{R. Jain\fnref{footnote7}}
\address{Physical Research Laboratory, Department of Space, Ahmedabad-380 009, India}
\fntext[footnote7]{Ph: +91-79-2631-4000, Fax: +91-79-2631-4900}
\ead{rajmal@prl.res.in}
\author{N.V. Nitta\fnref{footnote5}}
\address{Lockheed Martin Solar and Astrophysics Laboratory, Palo Alto, CA 94304, USA}
\ead{nitta@lmsal.com}
\author{H. Xie\fnref{footnote3}}
\address{NASA Goddard Space Flight Center, Greenbelt, MD 20771, USA}
\ead{hong.xie-1@nasa.gov}
\author{S. Yashiro\fnref{footnote3}}
\address{NASA Goddard Space Flight Center, Greenbelt, MD 20771, USA}
\ead{seiji.yashiro@nasa.gov}
\author{S. Akiyama\fnref{footnote3}}
\address{NASA Goddard Space Flight Center, Greenbelt, MD 20771, USA}
\ead{sachiko.akiyama@nasa.gov}
\author{P. M{\"a}kel{\"a}\fnref{footnote3}}
\address{NASA Goddard Space Flight Center, Greenbelt, MD 20771, USA}
\ead{pertti.makela@nasa.gov}
\author{P. Kayshap\fnref{footnote1}}
\address{Aryabhatta Research Institute of Observational Sciences (ARIES), Manora Peak, Nainital-263 129, India}
\ead{pradeep.kashyap@aries.res.in}
\author{A.K. Awasthi\fnref{footnote7}}
\address{Physical Research Laboratory, Department of Space, Ahmedabad-380 009, India}
\ead{awasthi@prl.res.in}
\author{V.C. Dwivedi\fnref{footnote4}}
\address{Radio Astronomy Centre, Tata Institute of Fundamental Research, Ooty-643 001, India}
\ead{vidya{\_}charan2000@yahoo.com}
\author{K. Mahalakshmi\fnref{footnote4}}
\address{Radio Astronomy Centre, Tata Institute of Fundamental Research, Ooty-643 001, India}
\ead{kmahee@gmail.com}
\begin{abstract}

We use multiwavelength data from space and ground based instruments to study the solar flares and coronal mass ejections (CMEs) on January 23, 2012 that were responsible for one of the largest solar energetic particle (SEP) events of solar cycle 24. The eruptions consisting of two fast CMEs ($\approx$1400 km s$^{-1}$ and $\approx$2000 km s$^{-1}$) and M-class flares that occurred in active region 11402 located at $\approx$N28 W36. The two CMEs occurred in quick successions, so they interacted very close to the Sun. The second CME caught up with the first one at a distance of $\approx$11-12 $R_{sun}$. The CME interaction may be responsible for the elevated SEP flux and significant changes in the intensity profile of the SEP event. The compound CME resulted in a double-dip moderate geomagnetic storm ($Dst\sim-73 nT$). The two dips are due to the southward component of the interplanetary magnetic field in the shock sheath and the ICME intervals.  One possible reason for the lack of a stronger geomagnetic storm may be that the ICME delivered a glancing blow to Earth.

\end{abstract}
\end{frontmatter}
\section{Introduction}
Solar flares and coronal mass ejections (CMEs) involve a sudden release of magnetic energy stored in complex active regions through magnetic reconnection \citep{Benz08}. The several mechanisms may compel the energy build-up in the flaring and eruptive regions that later released in form of bulk mass motion, heating, as well as acceleration of the energetic particles. These processes may involve magnetic instabilities, flux and helicity emergence, building of the magnetic field complexity, etc. \citep[and references cited therein]{Chand10,Sriva10,Shib11}. The two main space weather consequences of CMEs are Solar Energetic Particle (SEP) events and Geomagnetic Storms (GMSs). 

Both flares and CMEs may contribute to SEP events. According to \citet{Rea99}, there are two types of SEP events, impulsive and gradual. Impulsive SEP events, as characterized by enhanced $^3$He, heavy ions and electrons and high charge states, are attributed to impulsive solar flares. The solar sources of impulsive SEP events are identified with coronal jets and narrow CMEs \citep{Nitta06,Wang06,Yash04}. Gradual SEP events, which are far more important in the context of space weather, are due to acceleration of particles at CME-driven shocks \citep{kah84, Gop09,Mano11}.

Fast moving shocks associated with CMEs accelerate SEPs in the interplanetary medium near the Sun \citep{Gop08,Gop04} and at 1 AU in Energetic Storm Particle (ESP) events \citep{Bry62, Makela11}. CME interaction has been reported to increase the SEP acceleration efficiency \citep{Gop02,Gop04}.  The interaction between CMEs results in a new CME with the magnetic field and plasma from both the CMEs. The process is described as "CME cannibalism" \citep{Gop01}. Recently, \citet{Liu12} have reported on the implications of CME-CME interaction for shock propagation, particle acceleration and space weather forecasting. Such interactions have been modeled by three-dimensional MHD simulation \citep{Shen11}. Some authors have made statistical argument questioning the importance of CME interaction for SEP production \citep{Rich03}. However, if we consider CMEs from the same active region, the physical interaction is inevitable, so there must be some influence of the CME interaction on the SEP intensity profile \citep {Gop04,Li05}.

Intense geomagnetic storms occur when the southward field in the interplanetary CMEs (ICMEs) reconnect with the Earth's magnetic field \citep{Tsu88, Gon94}. Intense storms can also occur when the sheath region between the ICME and the shock has southward component of the magnetic field. It has been found that nearly 90\% of intense storms (Dst $\lesssim$ -100 nT) result from Earth-directed halo CMEs \citep{Zha07,Joshi11}.

In this paper we study multiple flares and CMEs from the same active region on January 23, 2012 that resulted in one of the largest SEP events of solar cycle 24. The sequence of eruptions also resulted in a moderate GMS. We study the interacting CMEs for the first time using multiple view points and extended field of view (FOV) available from the Solar Terrestrial Relations Observatory (STEREO) and Solar and Heliospheric Observatory (SoHO). In Sec.~2, we describe the data set used. We report the observational results in Sec.~3. In the last section (Sec.~4), we summarize the results.

\section{Data Sets}

We use imaging observations from the STEREO SECCHI suites \citep{How08}, including coronagraphs COR1 and COR2, and the Extreme Ultraviolet Imager (EUVI, \cite{Wue04}), the Large Angle and Spectrometric Coronagraph (LASCO, \cite{Bur95}) on board SoHO, the Atmospheric Imaging Assembly (AIA, \cite{Lem12}) and the Helioseismic Magnetic Imager (HMI, \cite{Sch12}) on board the Solar Dynamics Observatory (SDO, \cite{Pesn12}).  Radio spectral observations from Wind/WAVES \citep{Bou95} are used to study interplanetary radio bursts. We use H$\alpha$ images from Aryabhatta Research Institute of Observational Sciences (ARIES) \citep{Uddin06} and Global Oscillation Network Group (GONG) for chromospheric flare information. For the radio emission analysis, the Nobeyama Radioheliograph (NoRH) images at 17 and 34 GHz have been used \citep{Naka94}. Radio spectral data from the Hiraiso radio spectrograph (HiRAS) are used for metric type II radio burst information. Interplanetary Scintillation (IPS) observations from the Ooty Radio Telescope (ORT), RAC, NCRA-TIFR has been used to study the interplanetary (IP) propagation of CMEs \citep{Mano10}. We use ACE and Wind spacecraft data available at OMNI web (http://omniweb.gsfc.nasa.gov) for solar wind (in-situ) plasma (temperature, density, speeds, plasma beta) and magnetic field information. Information on CMEs observed by LASCO is basically taken from the online CME catalog (http://cdaw.gsfc.nasa.gov, \cite{Gop09}).  Soft X-ray flare and SEP information is obtained from Geostationary Operational Environmental Satellite (GOES). The Dst (in nT) index has been obtained from the World Data Center for Geomagnetism, Kyoto University, Japan (e.g., http://wdc.kugi.kyoto-u.ac.jp/dstdir/).


\section{Analysis and Results}  
On January 23, 2012 three solar flares (C2.5, M1.1 and M8.7) occurred, and the last two were associated with two major CMEs. The time line of the whole event is presented in Table~1. The upper panel of Fig.~1 shows the white light image and magnetogram of active region AR 11402, which is the source region of the eruptions. The active region was located at $\approx$ N28 W36, and had a simple $\beta$ magnetic field configuration.
The bottom panel of Fig.~1 shows the GOES soft X-ray plot in the $1.0-8.0$ \AA~and $0.5-4.0$ \AA~wavelength bands during 00:00 UT to 09:30 UT. The soft X-ray flux shows three flares (indicated by the arrows): first flare (C2.5) peaked around 01:44 UT, the second flare (M1.1) around 03:13 UT and the third one (M8.7) around 03:59 UT. Since all three flares occurred in quick succession, it was not possible to determine the exact onset time of the M1.1 and M8.7 flares. For the onset of the M1.1 flare we have used the time of the dip between the C2.5 and M1.1 flares. For the onset time of the M8.7 flare we have used the time when the d$I$/d$t$ turns positive (I is the soft X-ray intensity of the flare). The M1.1 and M8.7 are each associated with a CME. The second CME is associated with large filament eruption. In the upcoming subsections we discuss the details on the flares, associated CMEs and their interplanetary consequences.
\subsection{Details of Flare Evolution}
Fig.~2 shows images during the M1.1 flare in SDO/AIA 304, 171 and 94 \AA~wavelengths. The 304 \AA~channel gives information about the upper transition region, while 171 and 94 \AA~provides information about the coronal region and the flaring region, respectively. The charterstic log temperature (in MK) for 304, 171 and 94 \AA~lines are around 4.7, 5.8 and 6.8. The three columns show the images during the rise, maximum and the decay phase respectively. It is clear from the observations that the flare starts with two bright sources in 304 \AA~wavelength; one lies near the sunspot and other away from the sunspot. Overplotted magnetogram (Fig.~2e) shows that the kernels lie on the opposite magnetic polarity regions. We have overplotted the microwave sources (i.e., 17 and 34 GHz) over the SDO/AIA 304 \AA~images (see Figs.~2a, 2b and 2c). These sources lie over the bright kernels of the flare, which are part of the flare ribbons. Post flare loops connecting the flare ribbons have been observed in AIA 171 and 94 \AA~images around 03:29 UT (Figs.~2f and 2i). The M1.1 flare was associated with the first CME. The CME has an extended hot core ($>$6.3 MK) which will be discussed later. 

The long duration M8.7 flare starts with the eruption of the northern part of the active region filament. Fig.~3(a-c) show the eruption of the filament in AIA 304 \AA~wavelength. Fig.~3(d-f) show the same filament in STEREO-A EUVI 304 \AA~ wavelength. We show in Fig.~3g a height-time measurement of the filament on AIA 304 \AA~images. The filament shows slow rise with an average speed of $\approx$6 km s$^{-1}$ between 03:00-03:31 UT followed by a fast rise with an asymptotic speed of $\approx$590 km s$^{-1}$ during 03:40-03:46 UT. The average speed calculated from images in Fig.~3(d-f) is $\sim$280 km s$^{-1}$, consistent with speed derived form SDO/AIA observations of the filament corrected for projection effects during 03:26 UT to 03:46 UT. Fig.~3g clearly shows that the rise of the filament triggers the soft X-ray flare.

Fig.~4 shows the evolution of the M8.7 flare in the SDO/AIA 304, 171 and 94 \AA~wavelengths. The first, second and third columns of this figure show the images from the rise, the maximum and the decay phases of flare respectively. The flare starts with two bright kernels. We have overplotted the 17 GHz and 34 GHz microwave flare sources over the AIA 304 \AA~ flare arcade during the rise, maximum and decay phases (Figs.~4a, 4b and 4c). 
The microwave emission is from a subset of the flare loops observed in AIA 304 \AA. The HMI magnetogram contours plotted over the SDO/AIA 171 \AA~image (Fig.~4d), show that the flare ribbons are located at the opposite magnetic polarity regions of the active region. The flare ribbons and loops can also be seen in GONG and ARIES H$\alpha$ images (Fig.~5). In summary, both the M1.1 and M8.7 flares show typical flare arcades. The filament eruption is clear in the M8.7 flare, whereas in the M1.1 flare there was a hot ejecta which became the CME core in coronagraphic images (see details below). 

\subsection{CME Propagation, Radio Observations and SEP Event}
The M1.1 and M8.7 flares were accompanied by CMEs observed by SoHO and STEREO coronagraphs. STEREO B and A were located at -114 degrees and +108 degrees, respectively relative to Earth. STEREO-A COR1 and COR2 observed the two CMEs above the north-east limb. We used STEREO A data for height-time measurements, since the CMEs were limb events in STEREO-A view (less projection effects). Fig.~6(a-c) show the two CMEs (CME1 and CME2) from STEREO-A COR1 and EUVI 195 \AA~at 03:05 UT, 03:45 UT and 03:50 UT. We were able to image the CMEs in their early phase because of the extended FOV of STEREO COR1 (1.4-4 R$_{sun}$). The height time plots of the CMEs obtained using STEREO-A COR1 and COR2 data give speeds of CME1 and CME2 as $\approx$1400 km s$^{-1}$ and $\approx$2000 km s$^{-1}$, respectively (Fig.~6g). The CMEs occurred in quick succession: CME2 entered into the aftermath of CME1 at the first appearance of CME2 (Fig.~6b). The CME interaction took place due to the difference in speed, their origin from the same active region and their propagation in roughly the same direction. The interaction completed at a height of 11-12 $R_{sun}$ at $\sim$ 04:48 UT (Fig. 6g).

A sequence of LASCO C2 and C3 difference images in Fig.~7, shows the further evolution of the two CMEs. CME1 appeared in the LASCO C2 FOV at $\sim$03:12 UT at a position angle of 329$^{\circ}$ with a width of 221$^{\circ}$. The CME1 had a linear speed of 685 km s$^{-1}$, which is much smaller than the speed estimated from the STEREO measurements at lower heights. This is likely to be due to projection effects because the solar source at N28 W36 in Earth view, whereas the source was at the limb in STEREO-A view. CME1 had an extended core (see Fig.~7b). In COR1 FOV the core was also seen as a bright feature, which can be traced in EUV images as a hot ejecta. This is somewhat unusual because normally the CME core is a cooler filament. Fig.~8a shows the hot ejecta (dotted line) in the AIA 94 \AA~image at 02:22 UT. The hot ejecta was surrounded by CME1 (solid line) as can be seen in the AIA 193 \AA~difference image. This can also be seen in the STEREO-A EUVI 195 \AA~image at 02:35:30 UT (Fig.~8b). 

CME2 appeared at a height of 2.01 $R_{sun}$ in the COR1 FOV at around 03:40 UT. At this time, the leading edge of the CME1 was at 5 $R_{sun}$. When CME2 appeared in the LASCO/C2 FOV at 04:00 UT its leading edge had already reached the core of CME1 (core1) (see Fig.~7b). 
The linear speed of CME2 was $\approx$2175 km s$^{-1}$ in the LASCO FOV, which is comparable to the STEREO-A speed (the speed difference is within measurement error). The LASCO CME also seems to be closer to the sky plane. Fig.~7(c-f) shows that the two CMEs merged at 04:12 UT and thereafter moved out as a single compound CME. This can also be seen in the three frames of STEREO-A COR2 shown in Fig.~6(d-f).

Even though a metric type II bursts was not reported in the Solar Geophysical Data our examination of the dynamic spectrum from the HiRAS radio spectrograph, shows a type II burst feature between 03:40 and 03:50 UT. Given the high speed of CME2 it is unusual that the type II feature is very weak. This may be due to the fact that CME2 is running into the moving material of CME1 making it difficult to form a strong shock. Furthermore, the starting frequency of the type II burst was at 200 MHz. Assuming that the radio emission is at harmonic of the local plasma frequency, we see that the 200 MHz plasma level is at an unusually large height of 2.0 $R_{sun}$. This may be due to the fact that the shock is formed in the body of CME1. The interplanetary counterpart of the type II burst can be seen in the Wind/WAVES dynamic spectrum shown in Fig.~9a. An intense type III burst, which is due to electrons accelerated at the flare site, can also be seen. 

The DH type II burst stats around $\sim$04:00 UT, which is roughly the time the leading edge of CME2 reaches the core of CME1 (Fig.~6 (d-f) and Fig.~7(b)). At this time the shock seems to have become very strong due to the decline of the Alfven speed upstream of the shock. The type II is observed as a broad band feature at lower frequencies (Fig.~9(a-b)). The same shock accelerated protons as indicated by the SEP onset at $\sim$04:20 UT, which is about 40 minutes after the metric type II start (see Fig.~9c). However, if we consider the fact that $>$100 MeV protons take about 15 minutes to reach Earth, the SEP release at the Sun is close to the onset of the DH type II bursts. As noted before CME2 has already interacting with the hot core of CME1. There is also a second increase in the SEP intensity around 05:30 UT, which is right after the formation of compound CME. It is possible that some particles were trapped in the interaction region and released after the interaction ends. In the $\ge$10 MeV channel, the SEP intensity reaches a plateau of about $\approx$3000 pfu by 12:00 UT on January 23, 2012 until the shock arrival at 1 AU (on January 24, 2012 at 14:33 UT) and causing an ESP event (see Fig.~9d). The ESP event attained a peak proton flux of $\approx$6263 pfu at 15:30 UT on January 24, 2012. The SEP event was the largest in solar cycle 24 \citep{Gop12} as of this writing. 

The formation of shock in the near-Sun region (as shown by the type II radio emission) and shock signatures observed at 1 AU imply that the  compound CME was able to drive a shock and accelerate particles in the entire Sun-Earth distance. From the onset of CME2 in COR1 FOV (i.e., 03:40 UT on January 23, 2012) to the shock arrival at 1 AU ($\sim$14:33 UT on January 24, 2012), we see that the shock transit time is $\approx$35 hour.
\subsection{Interplanetary Evolution of the Compound CME}
Fig.~10 shows interplanetary scintillation (IPS) data of the CME event on January 23, 2012 observed by the ORT. The IPS technique provides information on the turbulent plasma at the front of the moving CME. The IPS observations describe the evolution of the compound CME beyond 50 R$_{sun}$. In the IPS FOV, between 100 and 225 $R_{sun}$, the CME decelerates from $\approx$900 to $\approx$700 km s$^{-1}$. The CME deceleration is mostly due to the interaction with the background solar wind flow \citep [and references therein] {Gop00, Vrsnak04, Mano06}. 

Fig.~11 shows the variation of interplanetary field parameters (total magnetic field (B), Z component of magnetic field (Bz), Electric field (Ey)), plasma parameters (solar wind speed (V), proton density (N), proton temperature (T), plasma beta ($\beta$)) and Dst index. The shock arrival is at $\approx$14:33 UT on January 24, 2012, which is followed by an extended sheath and a narrow interval of ICME material. The shock speed ($\sim$700 km s$^{-1}$) is consistent with the speed derived from IPS observations. The plasma beta and magnetic field plots do not show the classical signatures of a magnetic cloud. The narrow ICME suggests that its impact on Earth is at an angle rather than direct.  The Dst index shows that the ICME resulted in a moderate GMS (Dst $\sim$ -73 nT). A peak negative Dst of -73 nT was reached at $\approx$11:00 UT on January 25, 2012 which corresponds to the main phase of the moderate GMS. Actually there are two dips in Dst indicating two different interactions with the Earth magnetic field. First is due to the interaction of the sheath region with negative Bz with Earth's magnetic field; the second one is due to the negative Bz within the ICME.

\section{Discussion and Summary}

In this paper we presented the case study of the January 23, 2012 solar eruptions from NOAA AR 11402 (N28 W36) that resulted in a large SEP event and a moderate GMS. The eruptions consisted of two fast interacting  CMEs associated with M1.1 and M8.7 flares. We have shown that the interaction of the fast primary CME with the preceding CME resulted  in the huge SEP event. The results presented in this study are consistent with the earlier suggestions that the interaction between CMEs has important implications for large SEP events \citep{Gop02}. 

The January 23, 2012 events were observed by the STEREO mission taking advantage of its extended FOV much closer to the Sun, capturing CME interaction in more detail. It is significant that the metric type II burst was very weak but the interplanetary type II burst was very intense.  This may be attributed to different combinations of CME speed, Alfven speed and the physical conditions of the ambient medium. The temporal coincidence of the CME interaction with the onset of the intense interplanetary type II burst and the large SEP event is significant and suggests the plausible increase in the efficiency of particle acceleration. 

One of the interesting features of these eruptions is the fact that the core of CME1 was unusually hot because it was observed only in the hot plasma images (94 \AA~and 193 \AA). The temperature exceeded 6 MK which is a few times the average temperature of the corona. The SEP event started when the shock of CME2 propagated through the hot ejecta. The particles in the hot plasma were already energized by virtue of the high temperature, which makes it easy for the shock to accelerate these particles.

The CMEs presented in this paper resulted in a huge SEP event but not a strong GMS. One of the possible causes for the moderate GMS may be that only a small section (about 6 hours) of the ICME interacted with Earth. Following the CME motion using movies of STEREO COR2 images confirms that most of the compound CME propagated above the ecliptic. This is also seen in the STEREO and LASCO images in Figs.~6 and 7.

In this paper we have described multi-wavelength observations of flares and CMEs that are important in understanding how solar eruptions cause space weather. Deeper insights into the link between the solar origin of CMEs and their interplanetary and geo-space consequences can be obtained by a detailed modeling of CME initiation and propagation through extreme ambient conditions.\\

The key results of this study can be summarized as follows:\\

1. This study demonstrated that interaction between fast CMEs from the same active region can occur very close to the Sun as revealed by STEREO observations.

2. The onset of the intense DH type II bursts and the large SEP event coincided with the arrival of CME2 at the hot core of CME1.

3. The compound CME decelerated significantly between the Sun and the Earth (the CME speed decreased from $\approx$2000 km s$^{-1}$ near the Sun to $\approx$700 km s$^{-1}$ at 1 AU).

4. The SEP intensity profile shows interesting variations corresponding to the time of the formation of the compound CME.  

5. Only a small section of the ICME arrived at Earth because the CME propagation was at an angle to the Sun-Earth line. Therefore the resulting GMS was of moderate intensity.

6. The GMS was of double-dip nature because the shock sheath and the ICME contained negative Bz values and caused the two dips in the Dst index. \\

\noindent {\bf Acknowledgments}

{We thank the anonymoud referees for their valuable comments and suggestions. We thank IUSSTF/JC-Solar Eruptive Phenomena/99-2010/2011-2012 project on "Multiwavelength Study of Solar Eruptive Phenomena  and their Interplanetary Responses" for its support to this study during our bilateral collaboration. We acknowledge the Wind/WAVES and SDO's AIA and HMI teams for providing their data. SOHO is a project of international cooperation between ESA and NASA. The STEREO Science Center made available the data used in this work. N.C.J thanks Aryabhatta Research Institute of Observational Sciences (ARIES), Nainital for providing Post Doctoral Grant. This work was partly supported by NASA LWS program. RC, AKS and WU acknowledge ISRO/RESPOND Project no. ISRO/RES/2/379/12-13.}

\clearpage
\begin{table}[h]
\small
\begin{center}
\end{center}
\caption{\bf Timeline of solar events and their associated interplanetary and geo-space events}
\medskip
\begin{tabular}{lll}
\hline
S.No. & Date (Time) & Observations \\
\hline
1. & 23/01/2012 (00:39, 01:44 UT) & Start and peak times of C2.5 flare\\
2. & 23/01/2012 ($\approx$01:48 UT) & Hot ejecta (core 1) of CME1 in SDO 94 \AA\\
3. & 23/01/2012 (02:03, 03:13 UT) & Start and peak times of M1.1 flare\\
4. & 23/01/2012 ($\approx$02:22 UT) & CME1 above the limb in SDO 193 \AA\\
5. & 23/01/2012 (02:30 UT) & First appearance of CME1 in COR1 FOV\\
6. & 23/01/2012 (03:00-03:46 UT) & Filament eruption associated with the M8.7 flare\\
7. & 23/01/2012 (03:12 UT) & First appearance of CME1 in LASCO C2\\
8. & 23/01/2012 (03:35, 03:59; 04:35 UT) & Start, peak and end times of M8.7 flare\\
9. & 23/01/2012 (03:40 UT) & First appearance of CME2 in COR1 FOV\\
10. & 23/01/2012 (03:40 UT) & Onset of metric type II\\
11. & 23/01/2012 (04:00 UT) & First appearance of CME2 in LASCO C2\\
12. & 23/01/2012 (04:20 UT) & Start time of SEP event \\
13. & 23/01/2012 ($\approx$04:24 UT) & Formation of compound CME due to the\\
 & & interaction of CME1 and CME2\\
14. & 23/01/2012 ($\approx$04:48 UT & End time of interaction of CMEs 1 and 2 \\ 
15. & 23/01/2012 (04:00 UT) & IP Type II start time and end time\\
  & to 24/01/2012 ($\approx$15:00 UT) & \\
16. & 24/01/2012 (14:33 UT) & Shock arrival at Earth (sudden commencement)\\
17. & 25/01/2012 (11:00 UT) & Time of minimum Dst (-73 nT)\\
& & i.e., moderate GMS\\
\hline
\end{tabular}
\end{table}
\begin{figure}
\begin{center}
\hbox{
\includegraphics[width=1\textwidth,clip=]{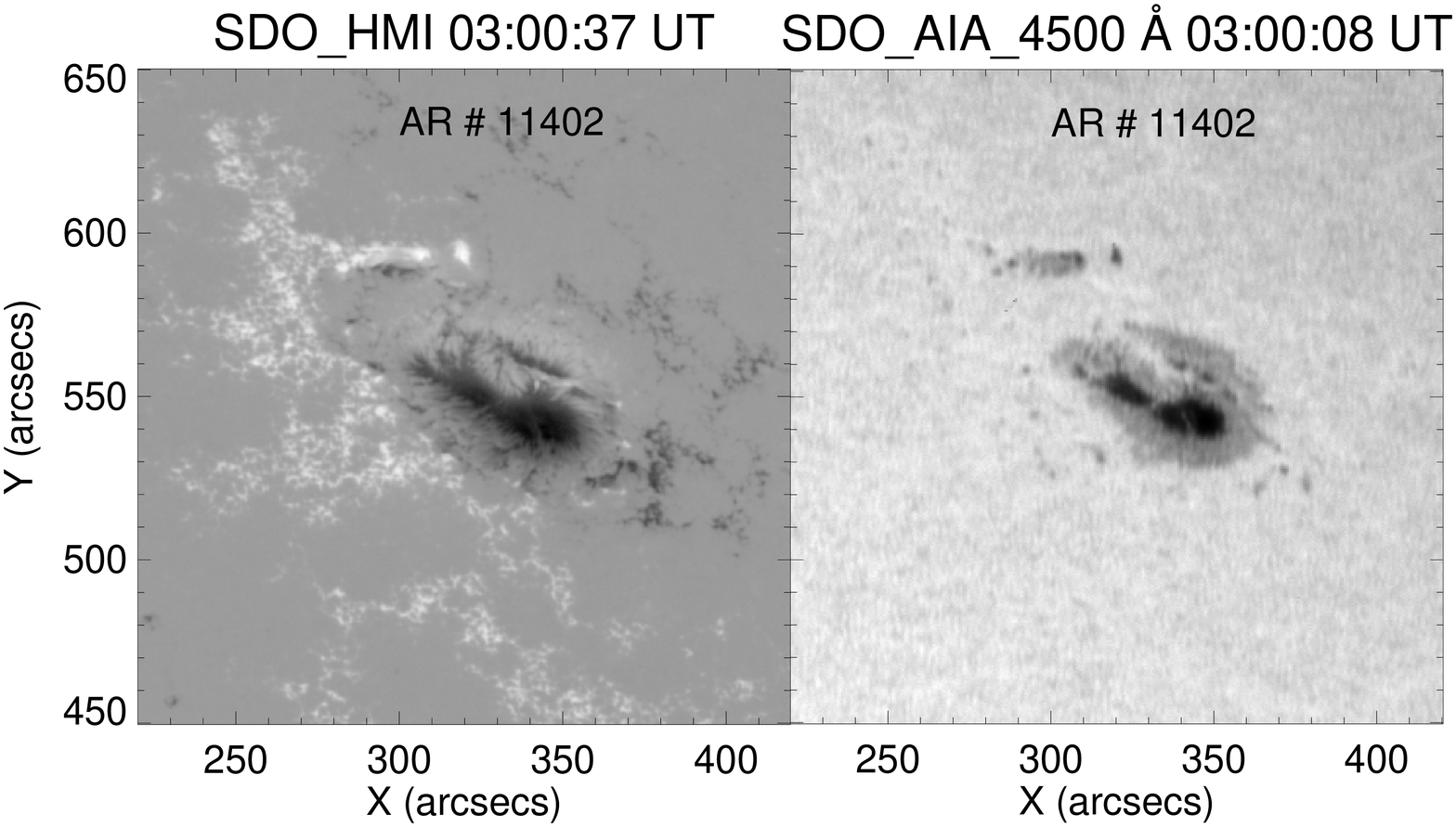}
}
\vspace*{-0.5cm}
\hbox{
\includegraphics[width=.9\textwidth,clip=]{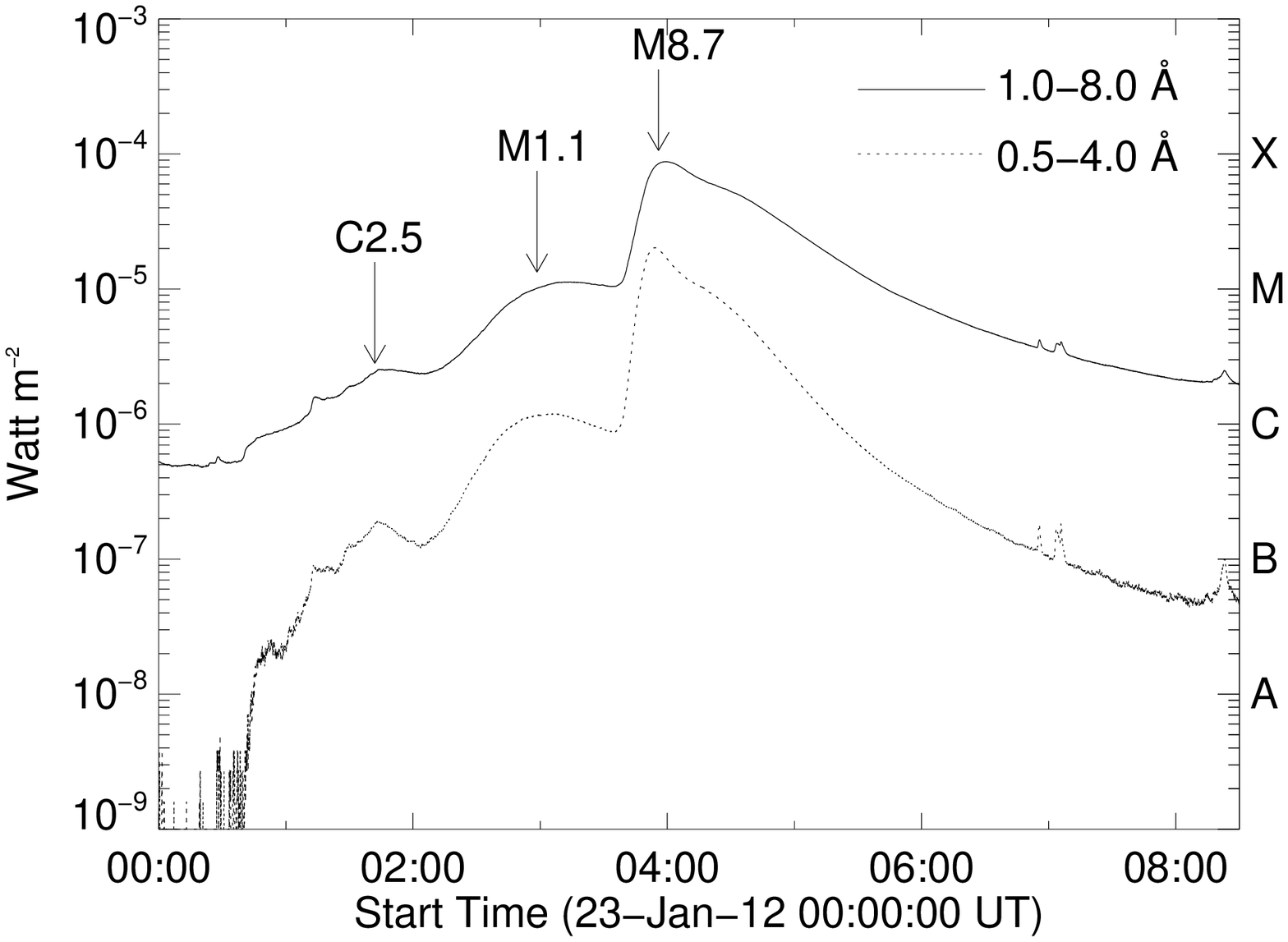}
}
\end{center}
\vspace*{-0.8cm} \caption{Upper panel: SDO/HMI and SDO/AIA 4500 \AA~images of the active region NOAA 11402 on January 23, 2012. Bottom panel: Temporal evolution of the GOES soft X-ray flux as observed by $GOES-15$ in the 0.5-4 \AA~and 1-8 \AA~bandwidths on 23 January, 2012.}
\label{}
\end{figure}
\begin{figure}
\begin{center}
\hbox{
\includegraphics[width=1\textwidth,clip=]{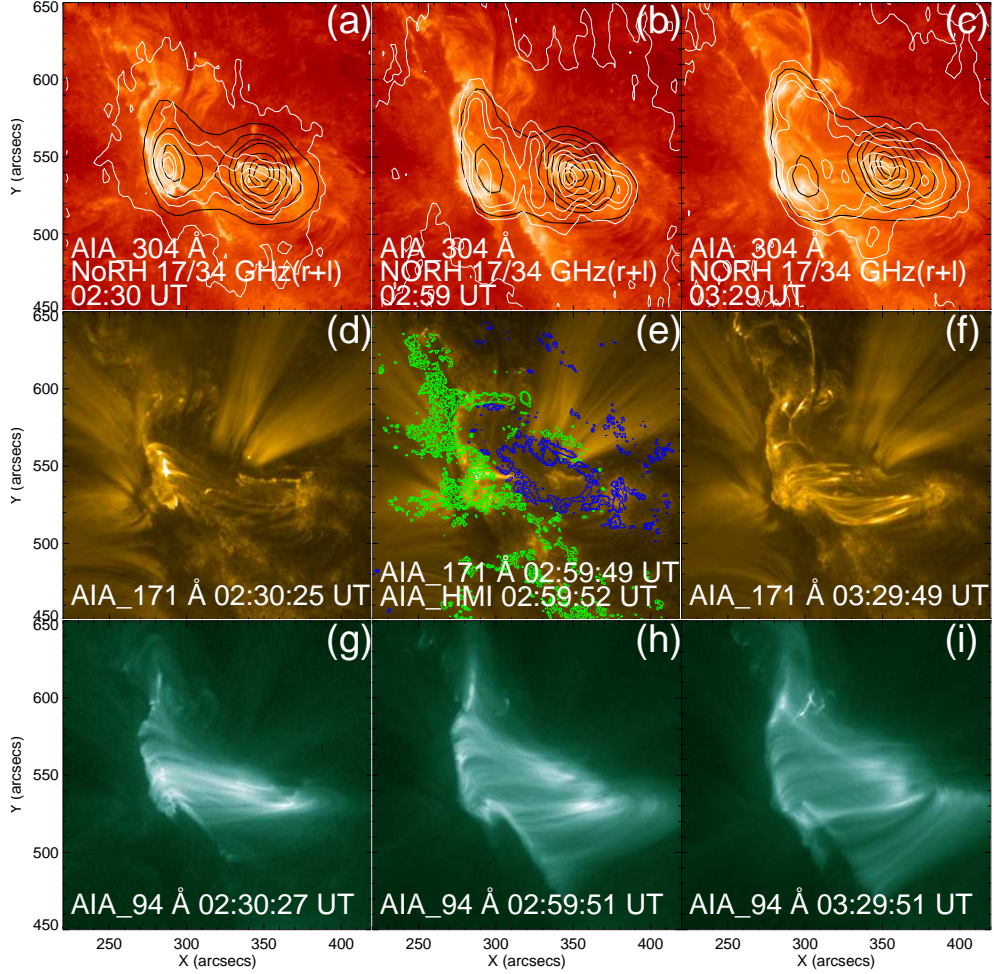}
}
\end{center}
\vspace*{-0.8cm} \caption{Evolution of M1.1 flare. Upper row: Evolution in 304 \AA~wavelength overplotted by microwave contours at 17 GHz (black contours) and 34 GHz (white contours). The contour levels are 10, 20, 30, 40, 60, 80, 90 percent of peak intensity. Middle row: Evolution of flare activity in 171 \AA~wavelength channel. The second image of this row (e) is overplotted by the SDO/HMI magnetogram contours. The green and blue contours show positive and negative polarity region respectively. The contour levels are $\pm{200}$, $\pm{400}$ Gauss. Bottom panel: Evolution of flare activity in 94 \AA~channel. The images in the first, second and third columns are corresponds to the rise, the maximum and the decay phases.}
\label{}
\end{figure}
\begin{figure}
\begin{center}
\vspace*{-3cm}
\hbox{
\hspace*{-3.6cm}
\includegraphics[width=.83\textwidth,clip=]{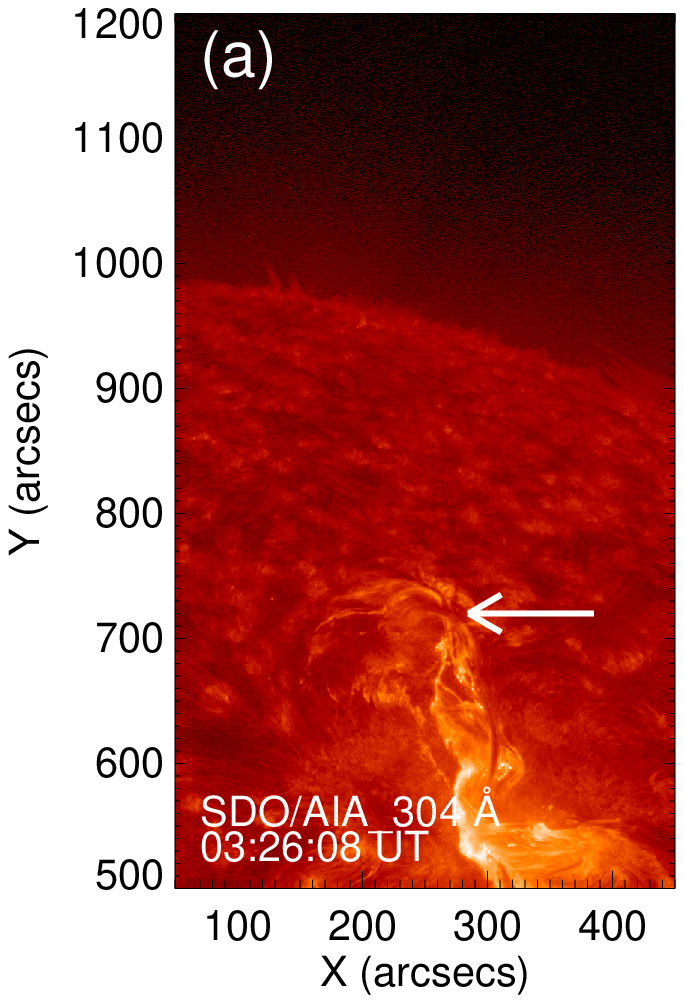}
\hspace*{-7cm}
\includegraphics[width=.83\textwidth,clip=]{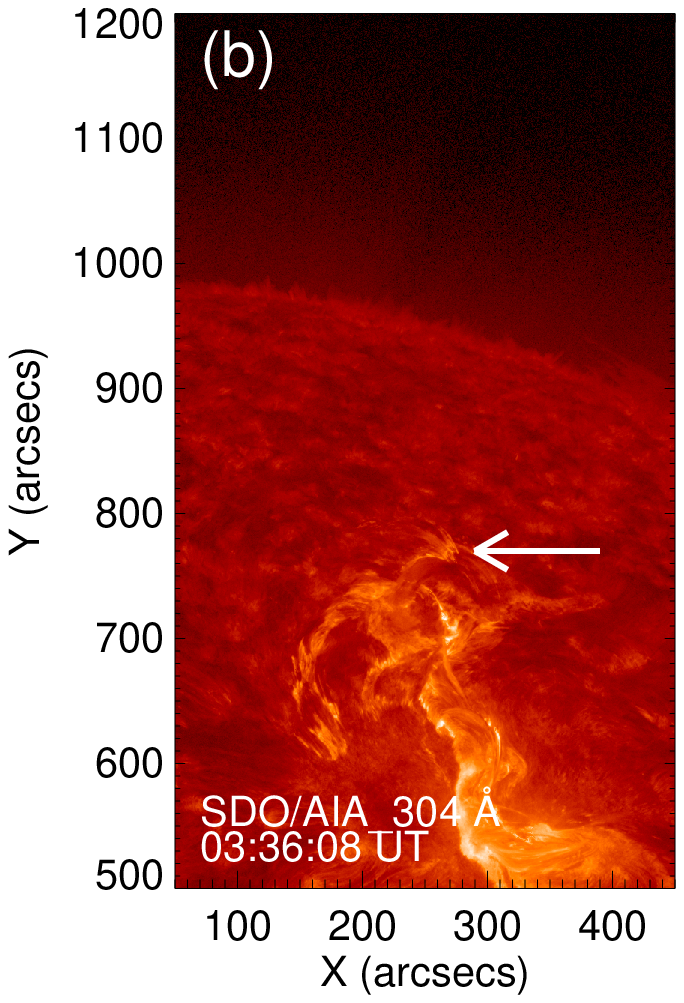}
\hspace*{-7cm}
\includegraphics[width=.83\textwidth,clip=]{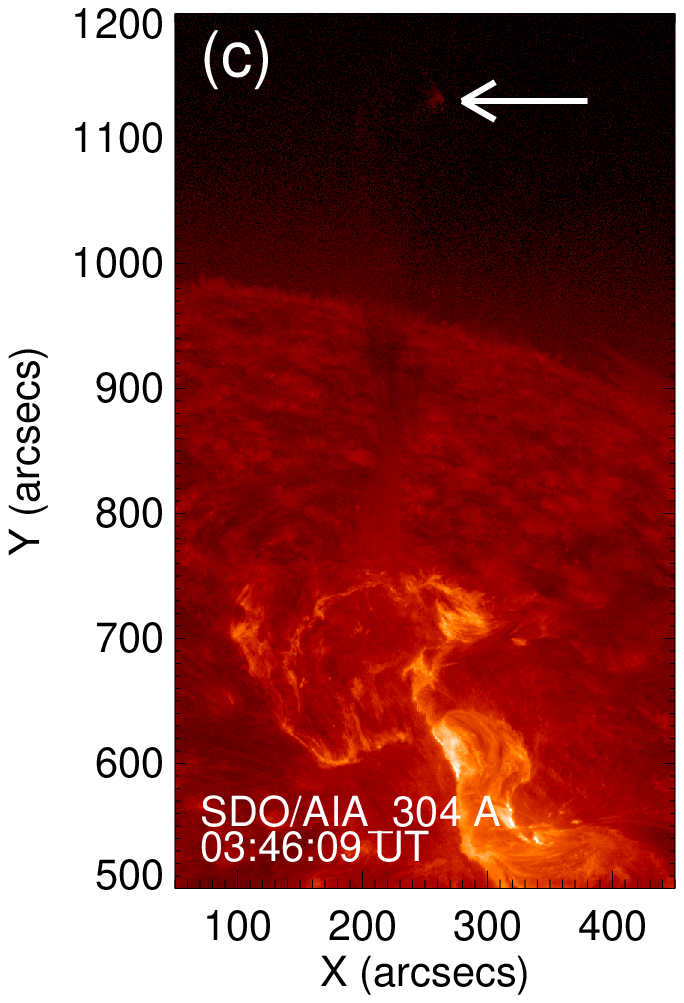}
}
\vspace*{-1.9cm}
\hbox{
\hspace*{-4.8cm}
\includegraphics[width=1\textwidth,clip=]{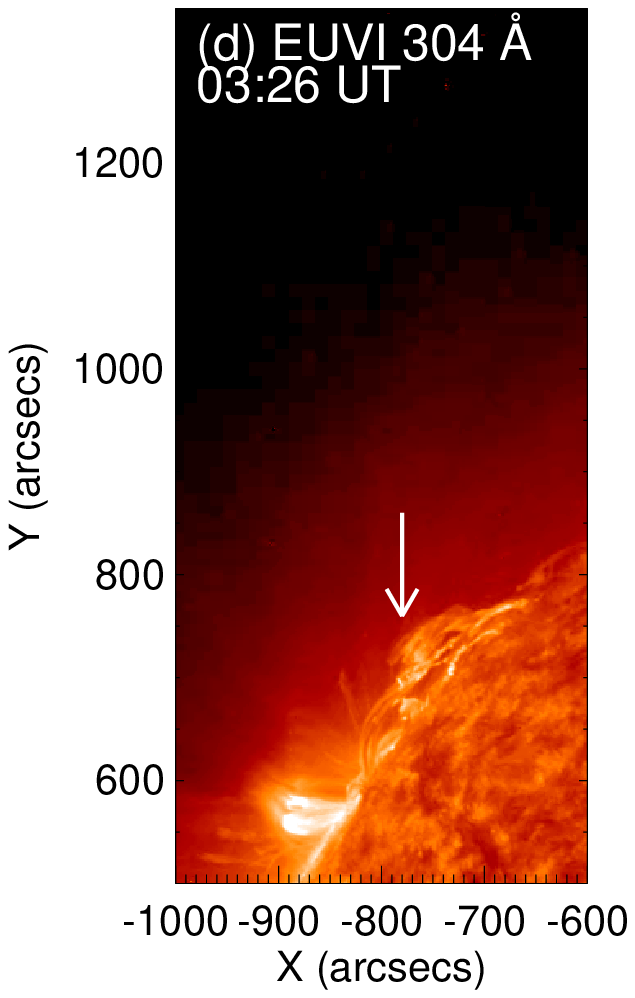}
\hspace*{-9.3cm}
\includegraphics[width=1\textwidth,clip=]{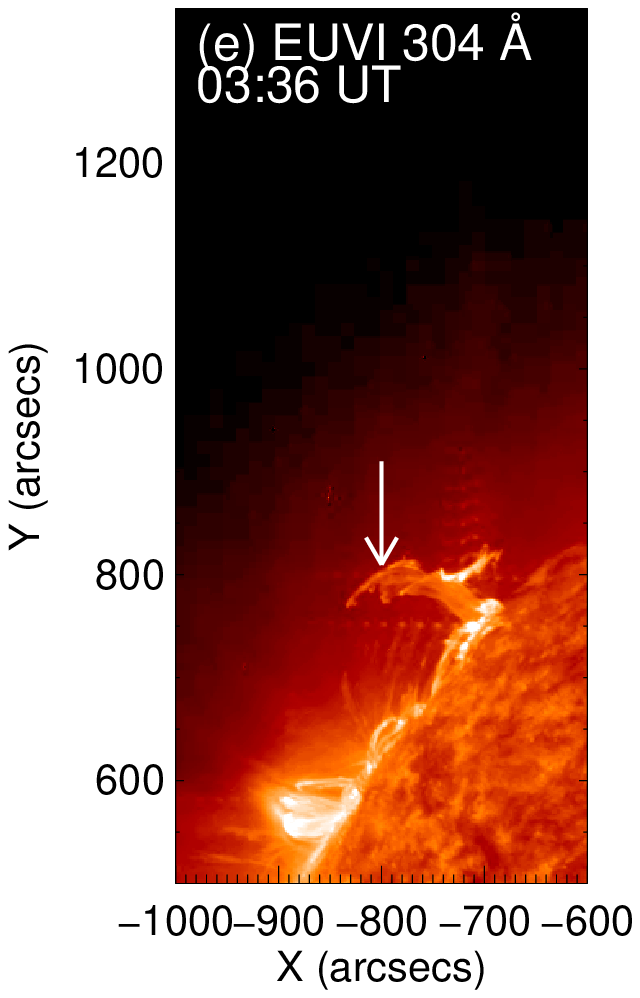}
\hspace*{-9.3cm}
\includegraphics[width=1\textwidth,clip=]{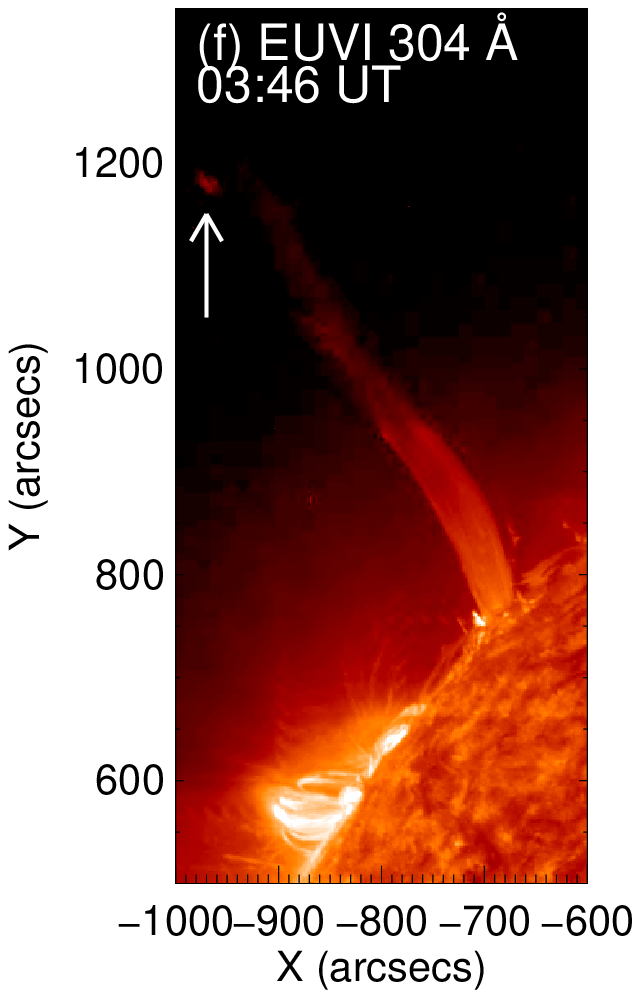}
}
\vspace*{-1cm}
\hspace*{0.45cm}
\hbox{
\includegraphics[width=.6\textwidth,clip=]{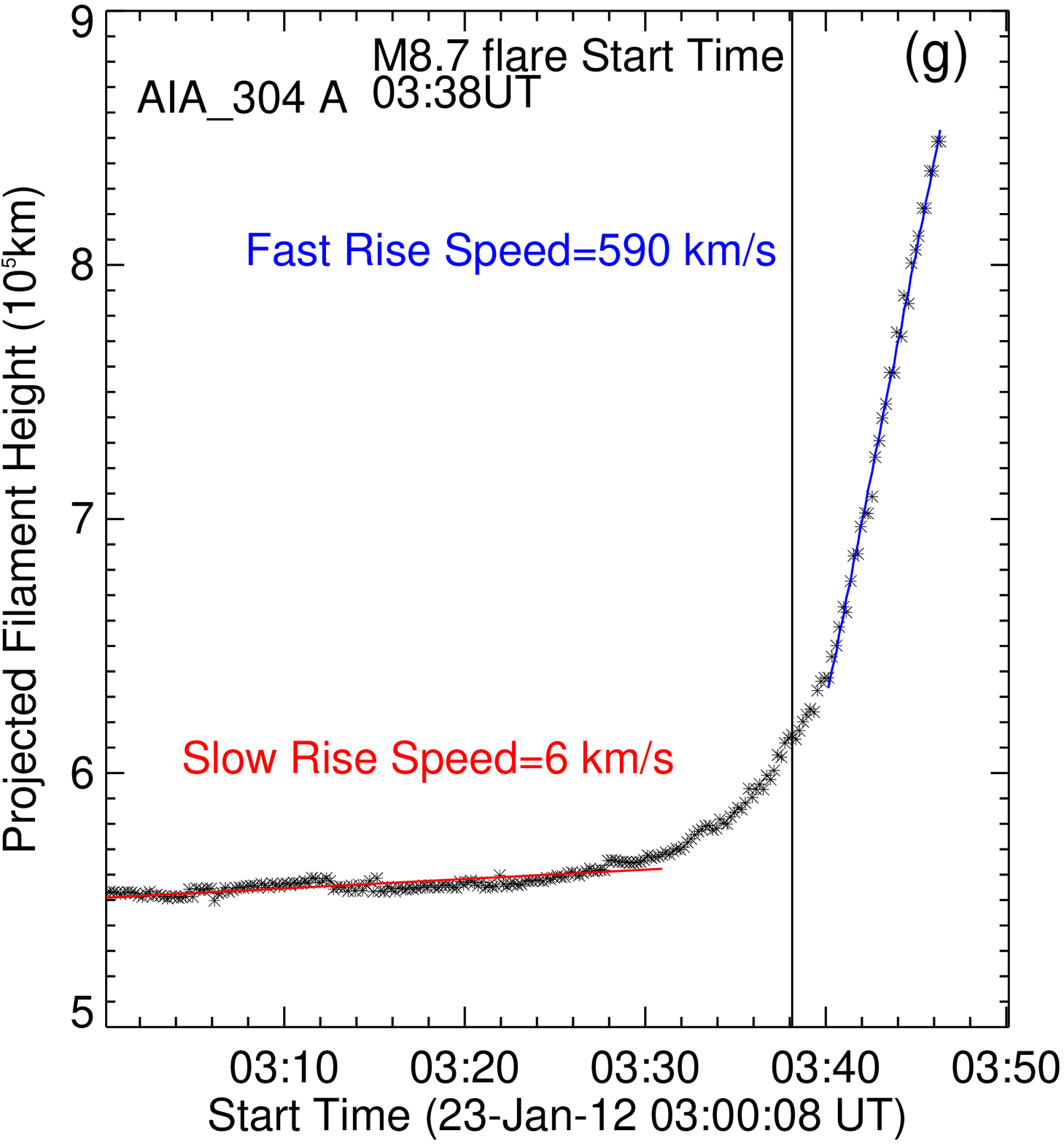}
}
\end{center}
\vspace*{-0.5cm} 
\caption{SDO/AIA 304 \AA~images (a-c) and STEREO-A EUVI 304 \AA~images (d-f) showing the evolution of the filament eruption. Arrows indicate the part of the filament where height-time measurements were made. Height-time plot of the filament obtained from AIA 304 \AA~images (g). The start time of the M8.7 flare is shown by the vertical line. A linear fit to the data points from 03:00 to 03:30 UT gives a speed of $\approx$6 km s$^{-1}$. A linear fit to the data points after the flare starts (from 03:40-03:46 UT) gives a speed of $\approx$590 km s$^{-1}$.}
\label{}
\end{figure}
\begin{figure}
\begin{center}
\hbox{
\includegraphics[width=1\textwidth,clip=]{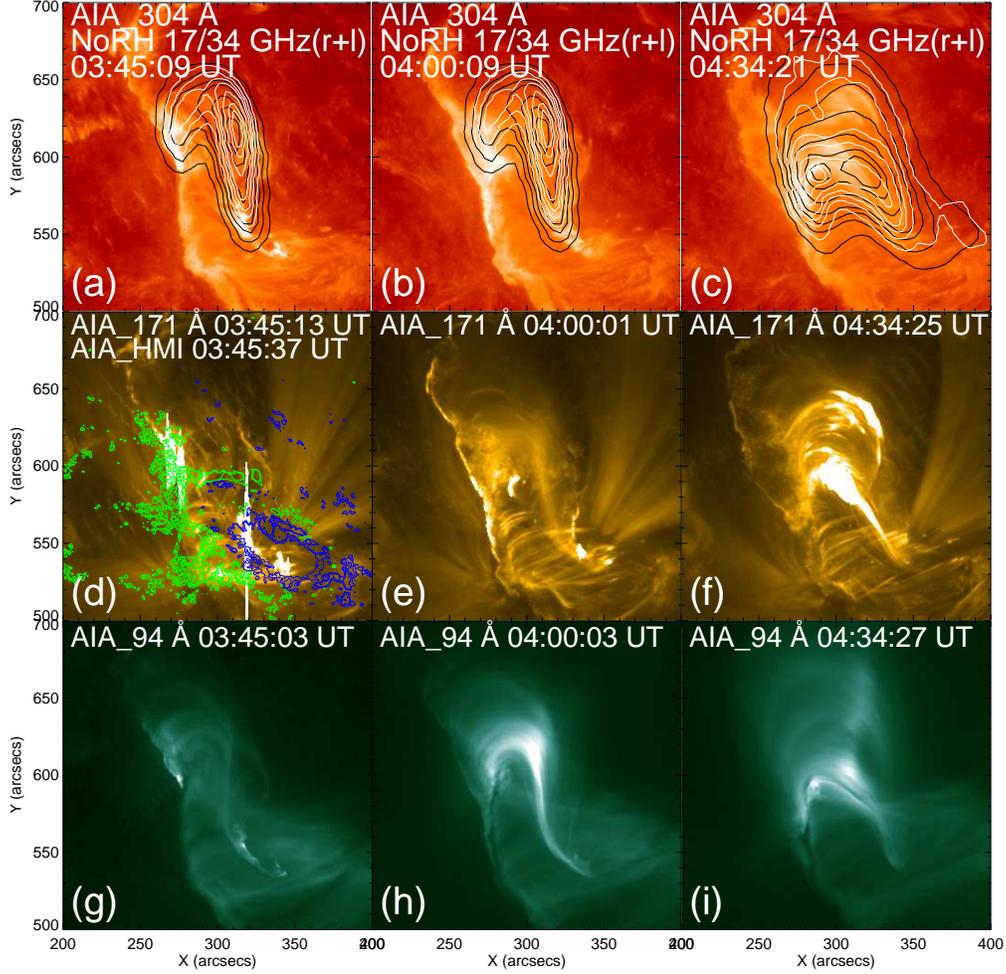}
}
\end{center}
\vspace*{-0.8cm} \caption{Evolution of M8.7 flare. Upper row: Evolution of the flare in 304 \AA~wavelength overplotted by microwave contours at 17 GHz (black contours) and 34 GHz (white contours). The contour levels are 10, 20, 30, 40, 60, 80, 90 percent of peak intensity. Middle row: Evolution of flare in 171 \AA~wavelength channel. The first image of this row (d) is overplotted by the magnetogram contours. The green and blue contours show positive and negative polarity region respectively. The contour levels are $\pm{200}$, $\pm{400}$ Gauss. Bottom panel: Bottom row: Evolution of flare in 94 \AA~channel. The images in the first, second and third columns are corresponds to the rise, the maximum and the decay phases.}
\label{}
\end{figure}
\begin{figure}
\begin{center}
\hbox{
\hspace*{-3cm}
\includegraphics[width=.8\textwidth,clip=]{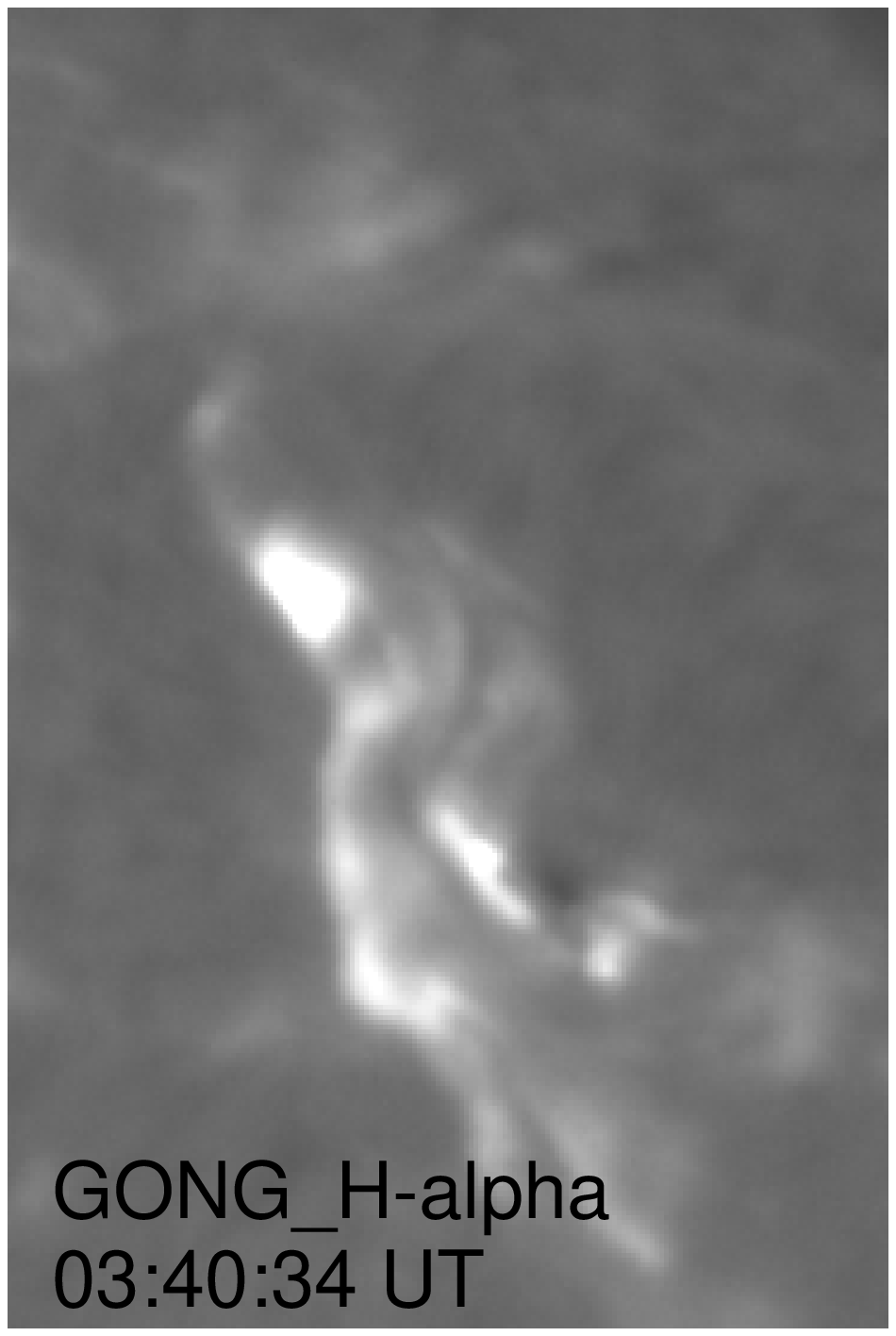}
\hspace*{-6cm}
\includegraphics[width=.8\textwidth,clip=]{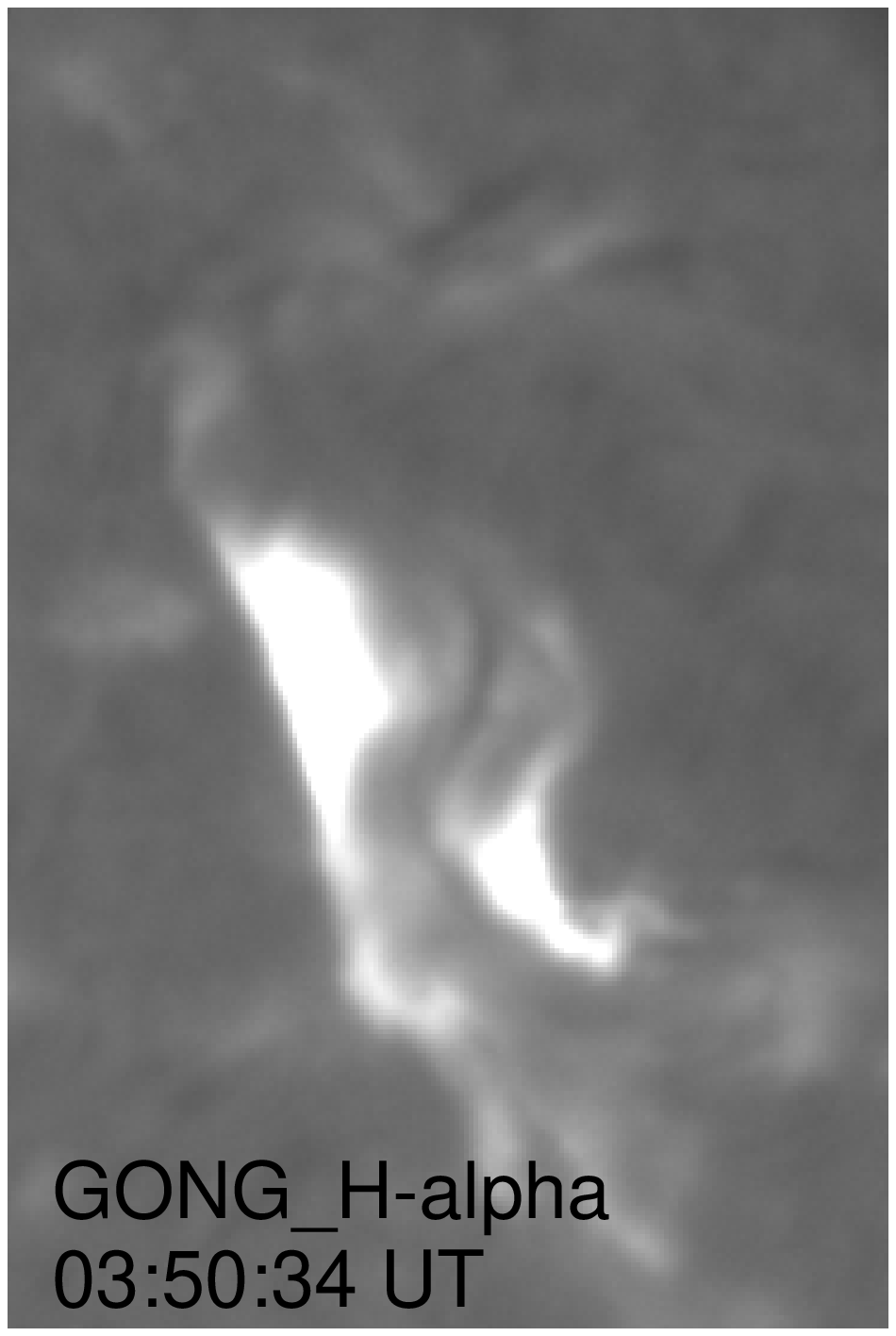}
\hspace*{-6cm}
\includegraphics[width=.8\textwidth,clip=]{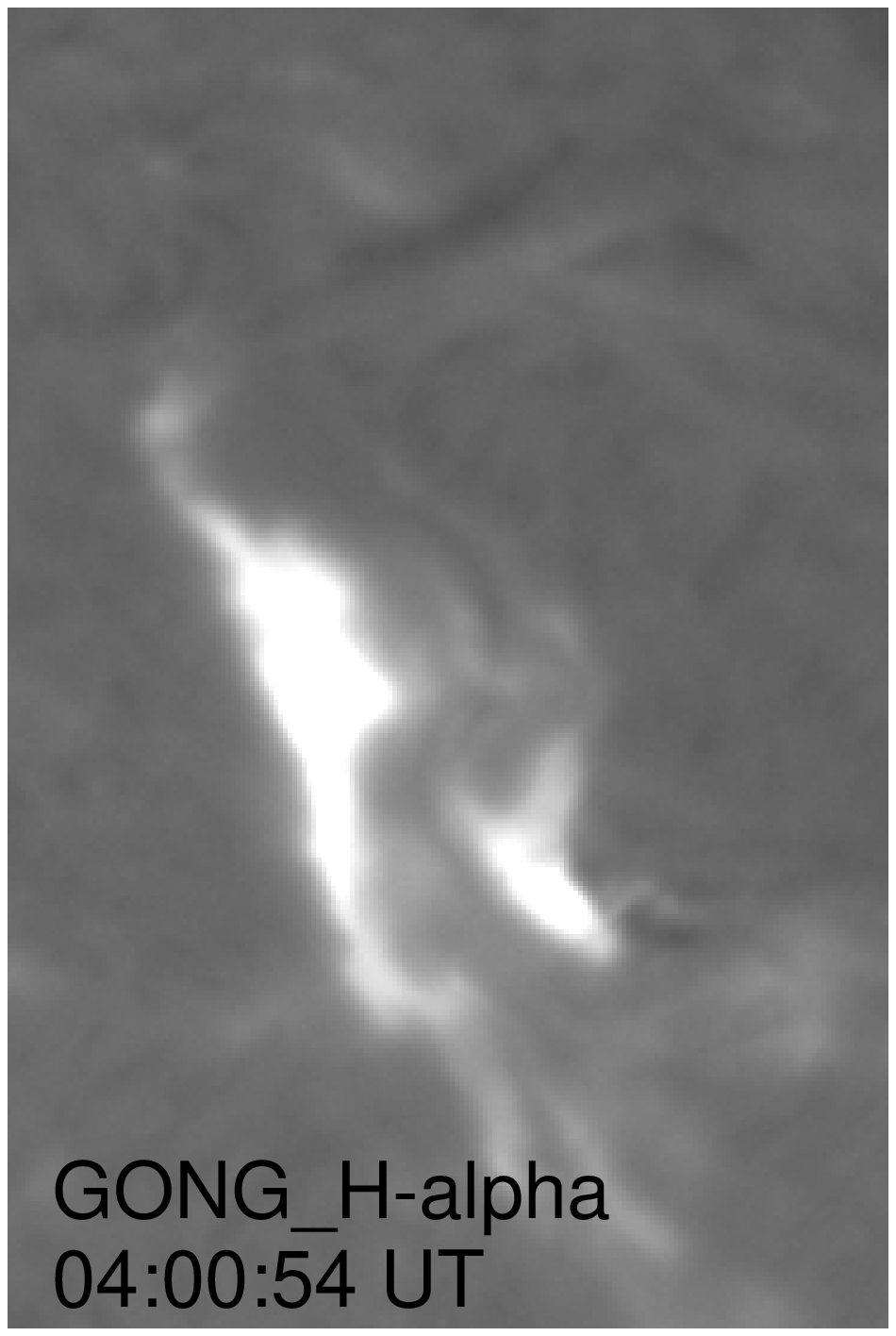}
}
\vspace*{-3cm}
\hbox{
\hspace*{-3cm}
\includegraphics[width=.8\textwidth,clip=]{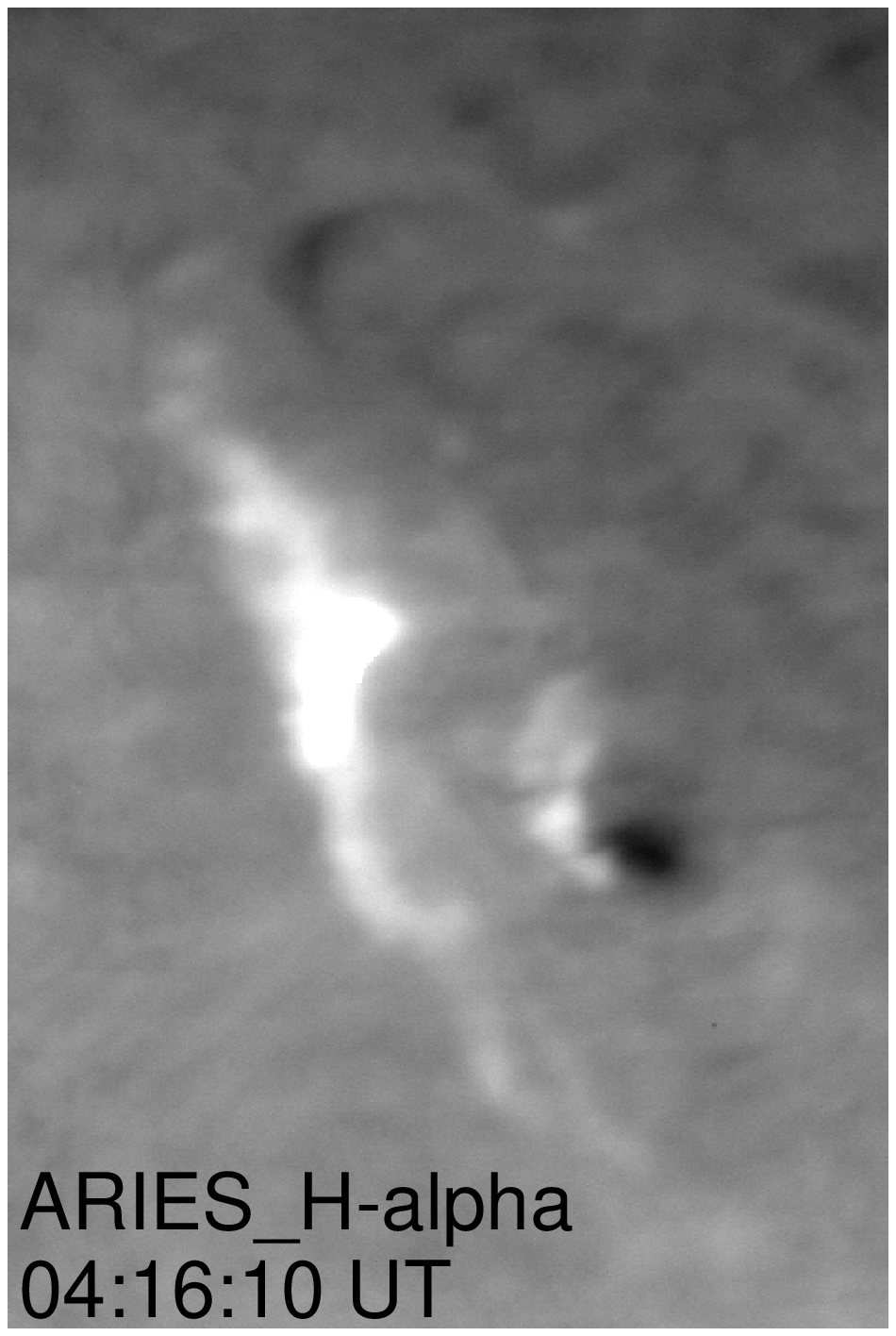}
\hspace*{-6cm}
\includegraphics[width=.8\textwidth,clip=]{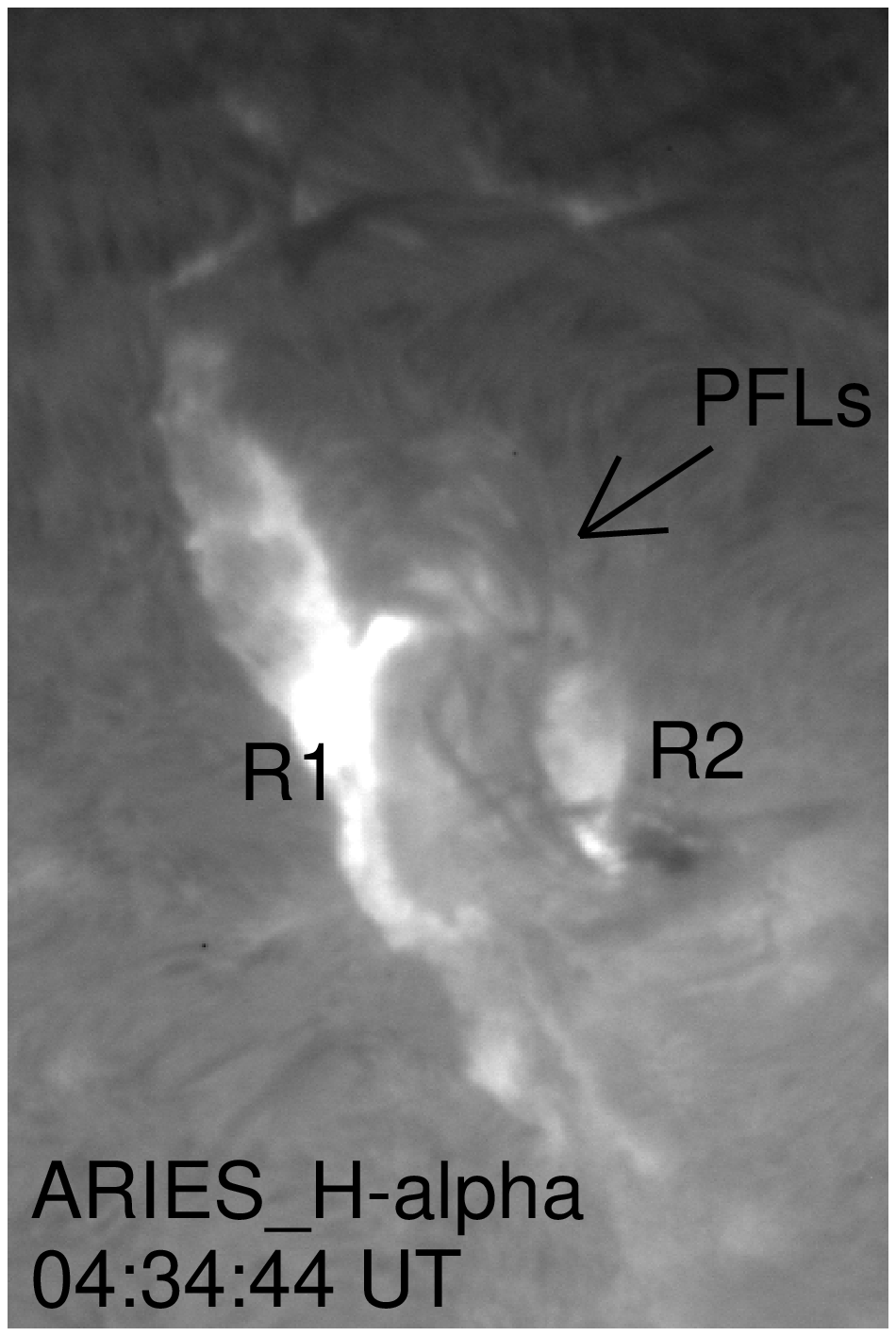}
\hspace*{-6cm}
\includegraphics[width=.8\textwidth,clip=]{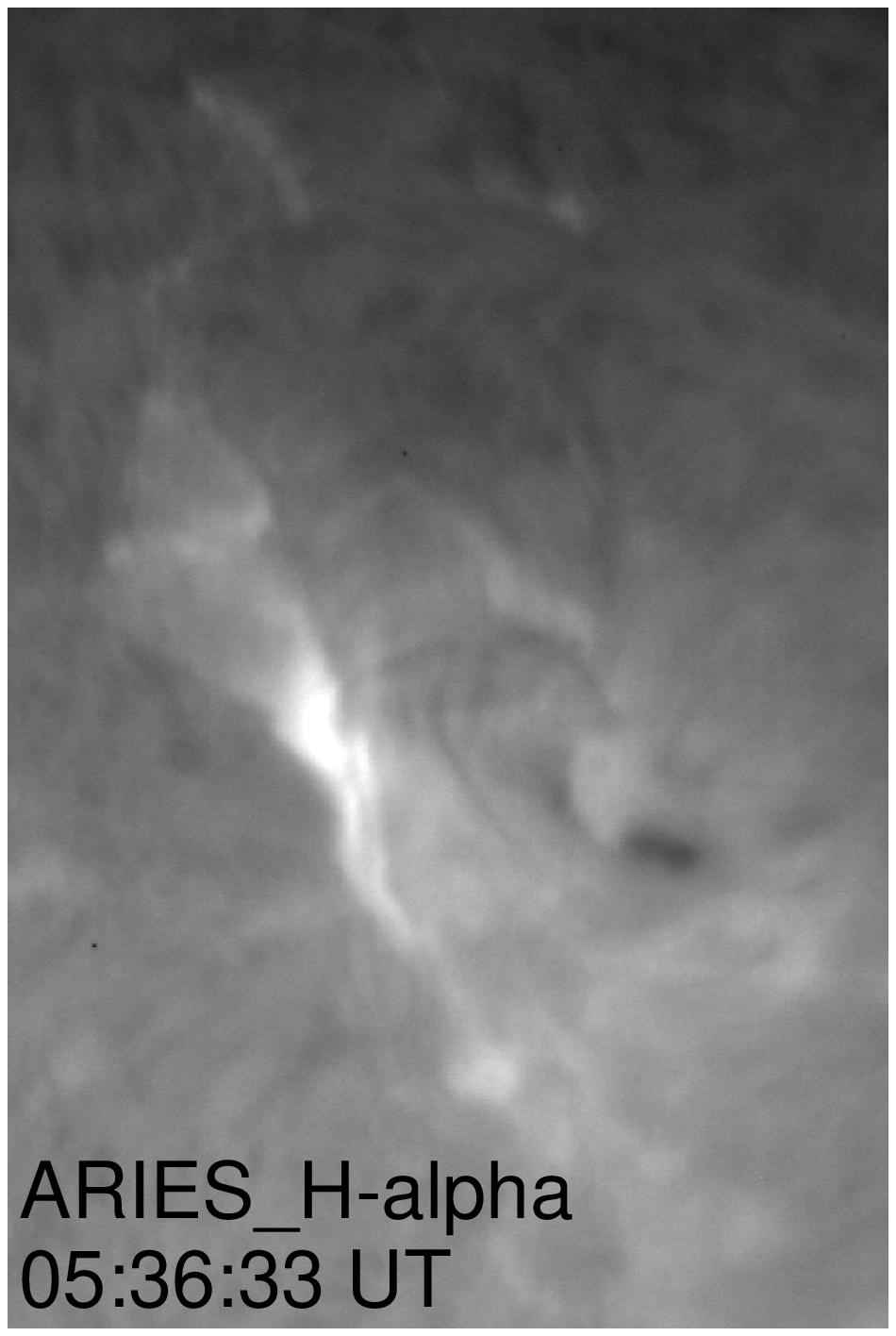}
}
\end{center}
\vspace*{-2cm} 
\caption{Upper panel: GONG H$\alpha$ images showing the rising phase of the M8.7 flare (FOV $200^{"} \times 300^{"}$). Bottom panel: ARIES H$\alpha$ images showing the decay phase dynamics of the M8.7 flare. 'R1' and 'R2' in the bottom middle figure indicates the eastern and western flare ribbon. Post Flare Loops (PFLs) are also indicated by the arrow in the middle panel at 04:34:44 UT. North is up and west is to the right (FOV $232^{"}\times 348^{"}$).}
\label{}
\end{figure}
\begin{figure}
\begin{center}
\hbox{
\includegraphics[width=1\textwidth,clip=]{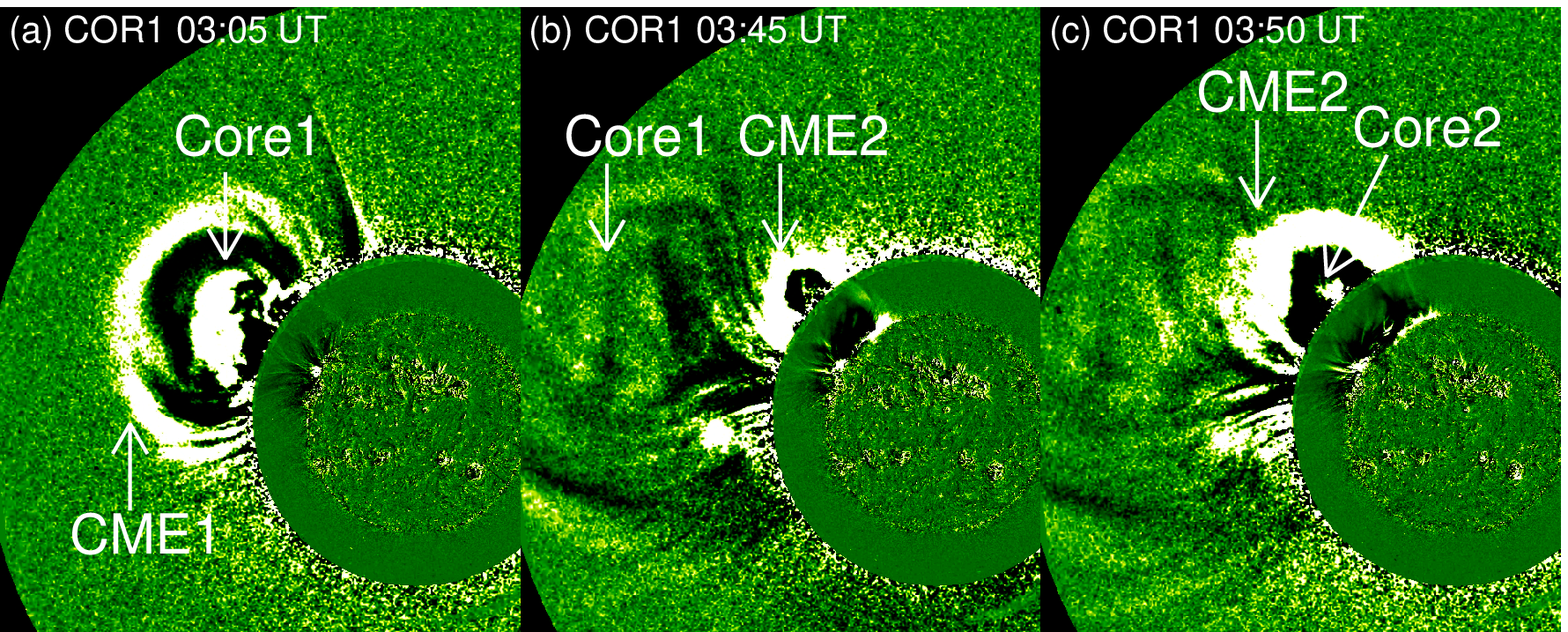}
}
\hbox{
\includegraphics[width=1\textwidth,clip=]{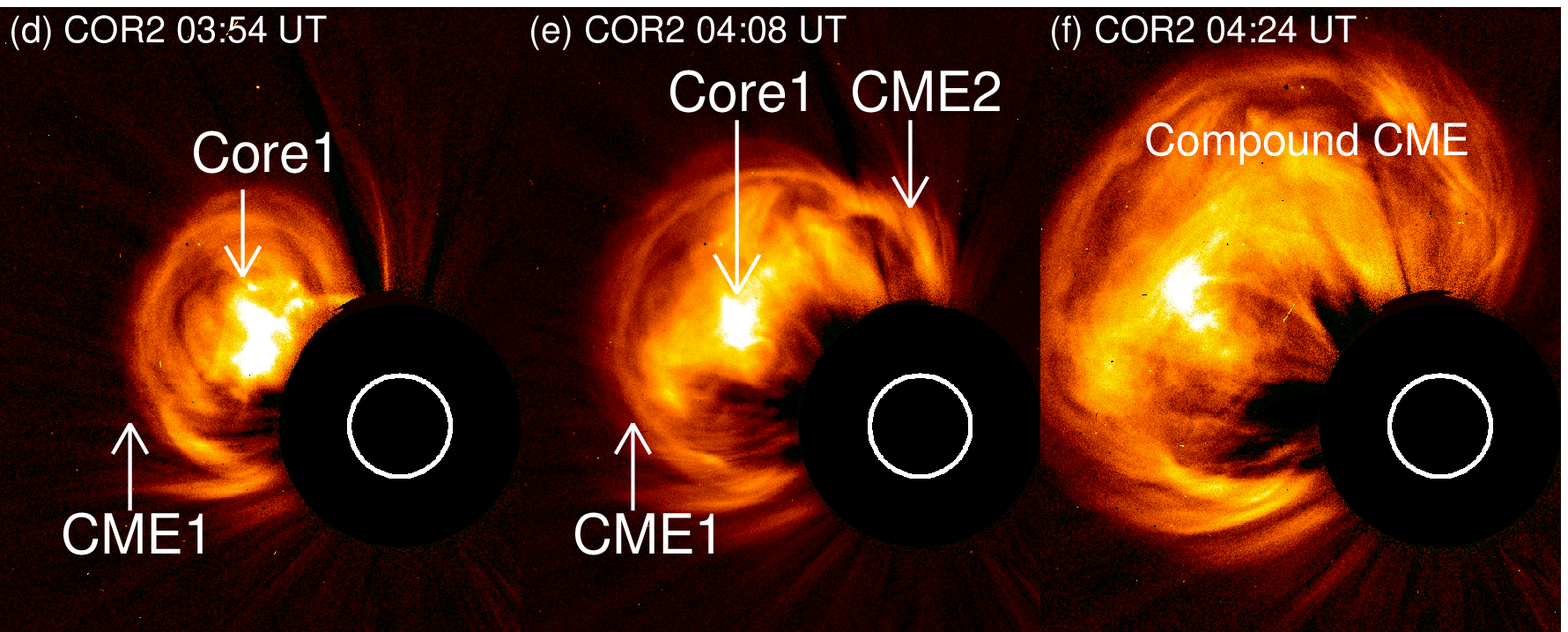}
}
\vspace*{-0.1cm}
\hbox{
\includegraphics[width=.8\textwidth,clip=]{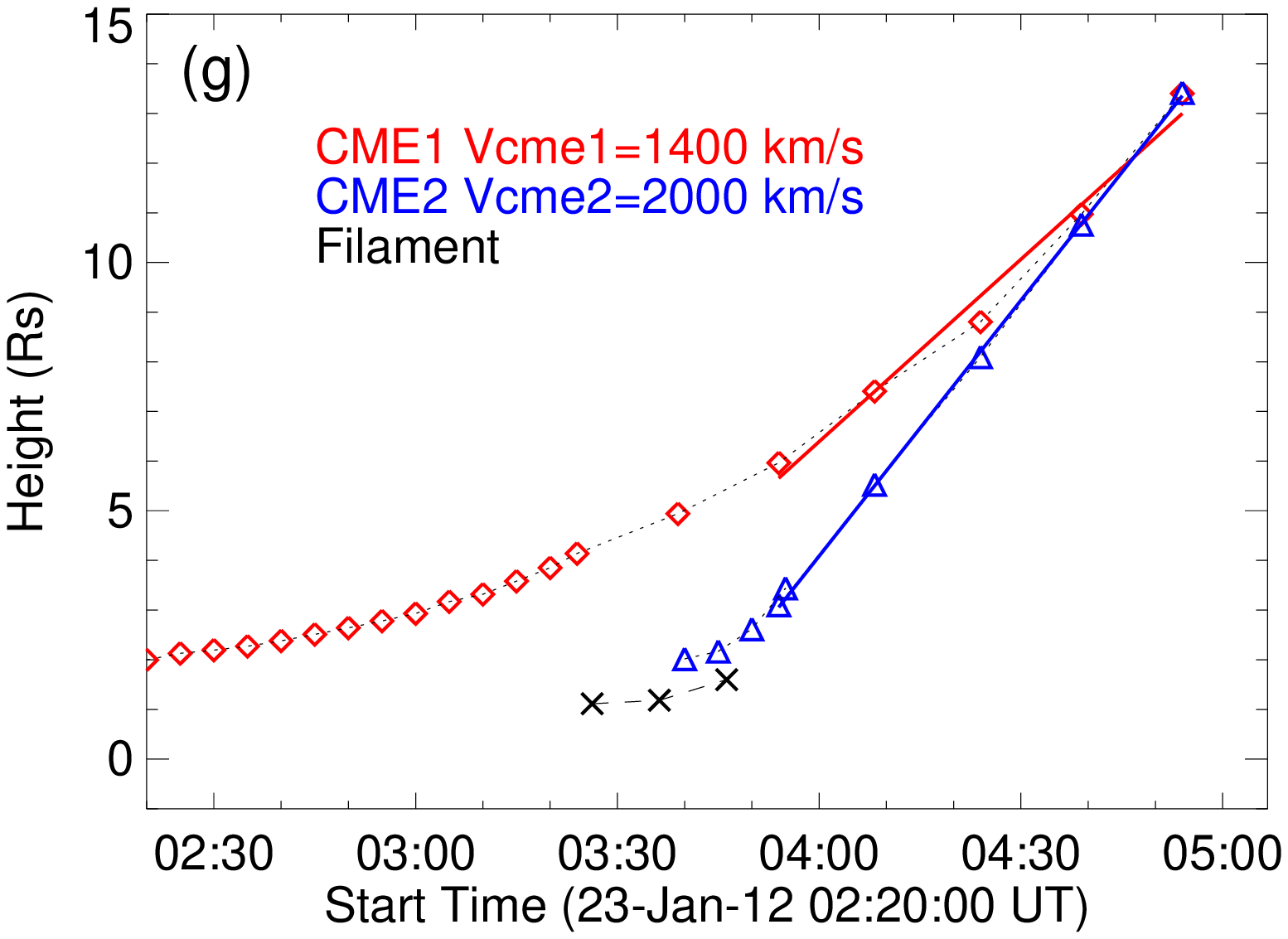}
}
\end{center}
\vspace*{-1cm} \caption{EUVI full disk running difference images superposed on the STEREO-A/COR1 difference images at 03:05 UT, 03:45 UT and 03:50 UT (a-c) showing two CMEs associated with the M1.1 and M8.7 flares respectively. STEREO COR2 base difference images at 03:54 UT, 04:08 UT and 04:24 UT (d-f) showing the CME1, CME2 and compound CME. Height-time plot of the CMEs using STEREO-A COR1 and COR2 (CME1: diamonds; CME2: triangles) (g). The speeds were computed by fitting a straight line to the last five data points. The position angle of our measurement was around $\approx$60$^{\circ}$. The dark dashed line with cross symbols represents the filament trajectory measured from STEREO EUVI 304 \AA~images.}
\label{}
\end{figure}

\begin{figure}
\begin{center}
\vspace*{-0.8cm}
\hbox{
\includegraphics[width=1\textwidth,clip=]{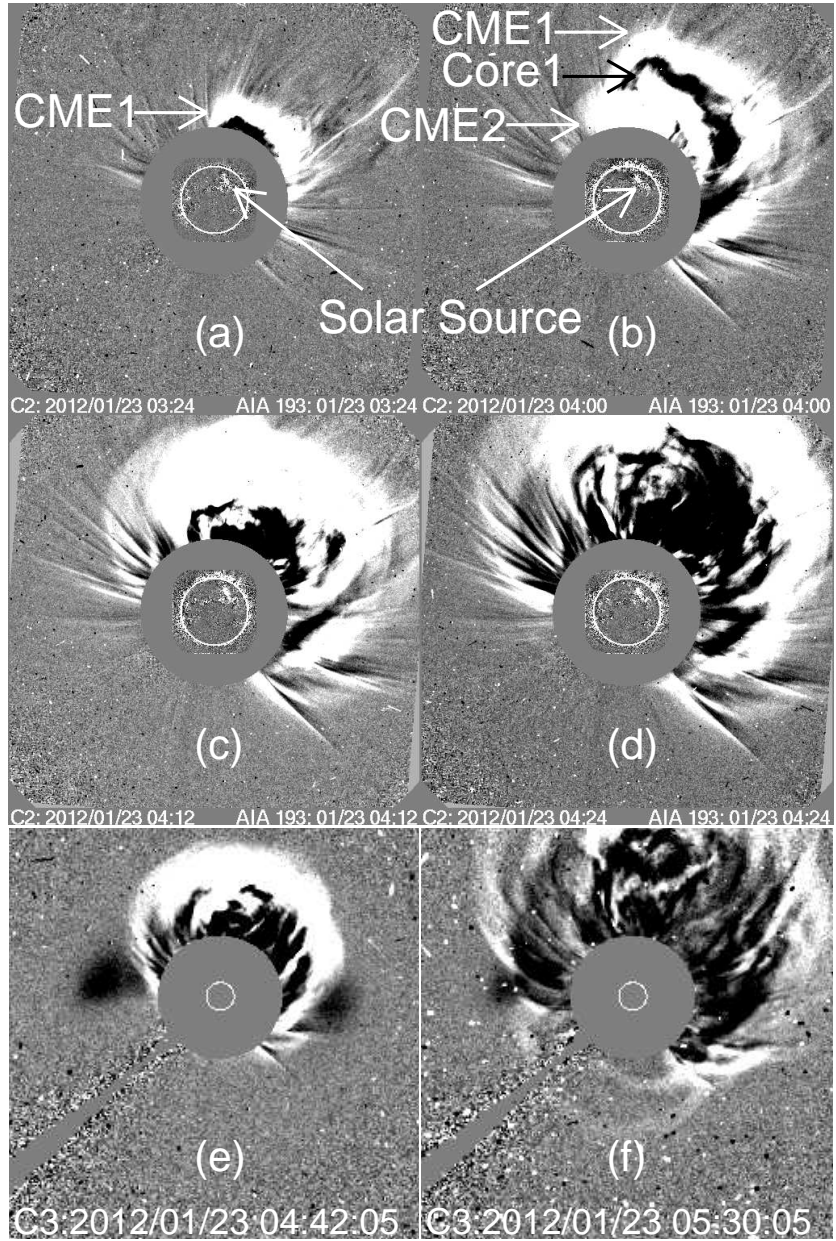}
}
\end{center}
\vspace*{-1cm} \caption{Difference images of LASCO C2 and C3 coronagraphs on-board SoHO showing the two CMEs (a-b) associated with M1.1 and M8.7 solar flares. CME-CME interaction can be seen in the middle row (c-d). Bottom panel shows the compound CME (e-f).}
\label{}
\end{figure}
\begin{figure}
\begin{center}
\vspace*{-2cm}
\hspace*{-4cm}
\hbox{
\includegraphics[width=1\textwidth,clip=]{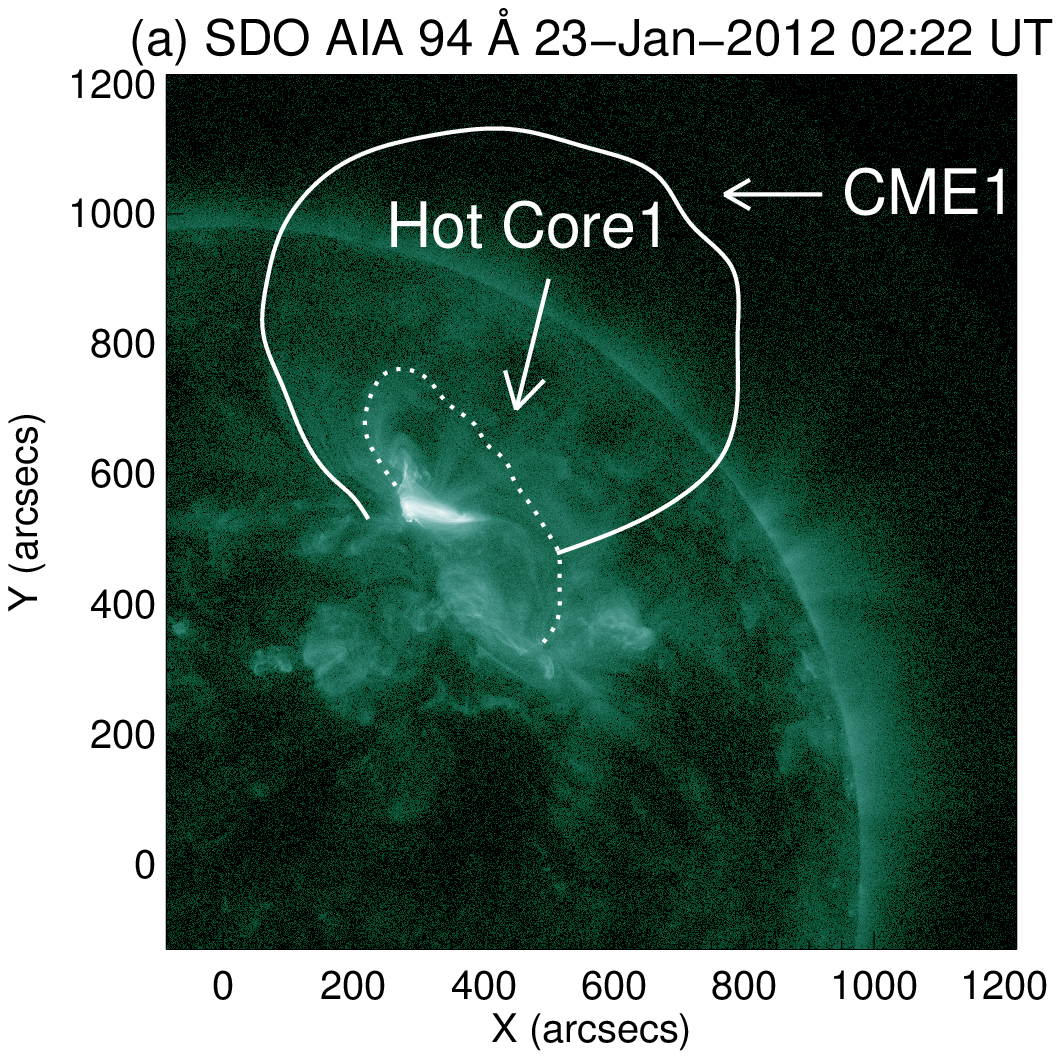}
\hspace*{-4cm}
\includegraphics[width=0.72\textwidth,clip=]{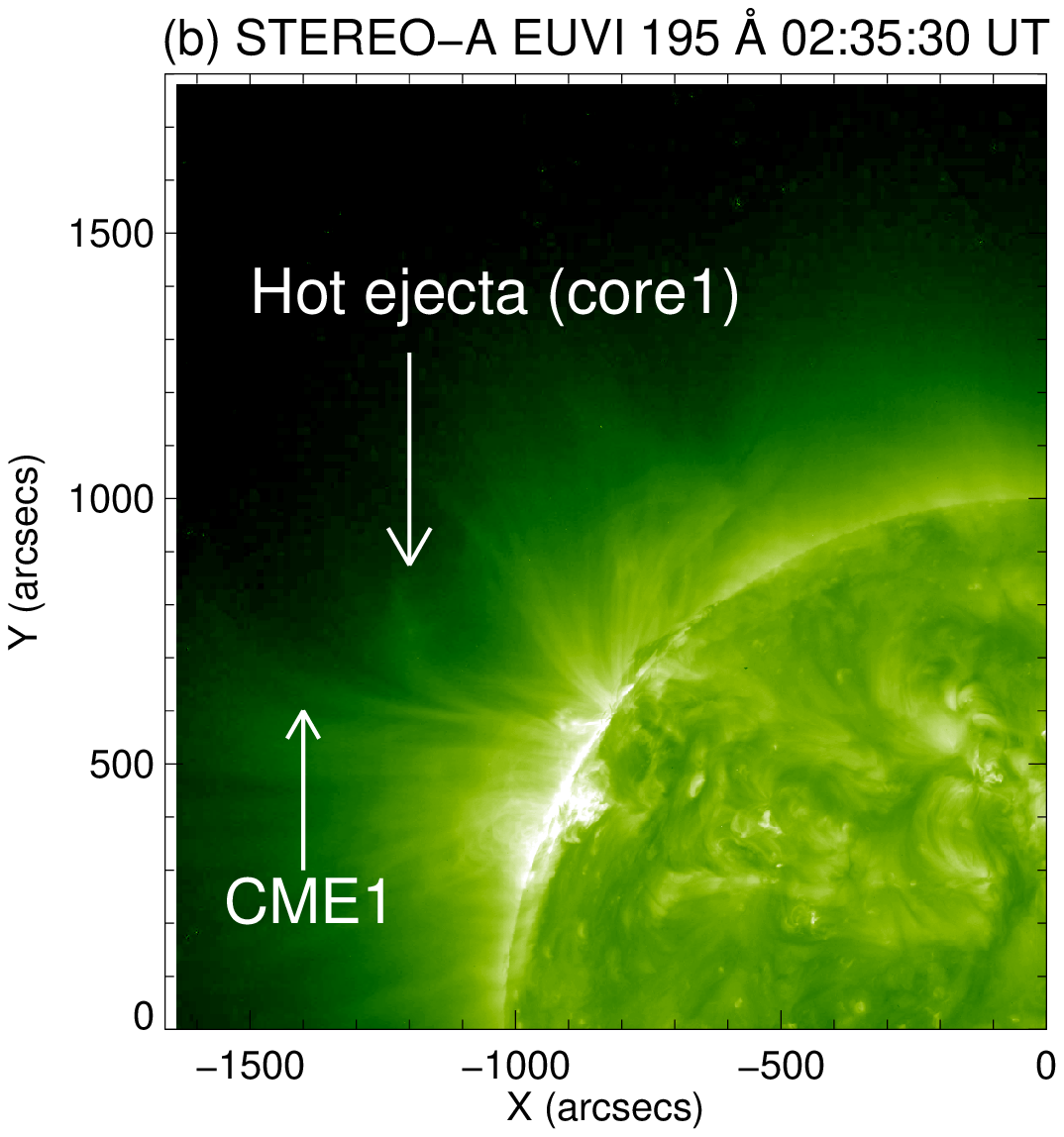}
}
\end{center}
\vspace*{-1cm} \caption{(a). AIA 94 \AA~ image showing the outer edge of hot ejecta and CME1 leading edge. Dotted curve showing the leading edge of the hot coronal loop and the solid curve shows the leading edge of the CME1. The CME1 leading edge points have been measured from the AIA 193 \AA~difference images. (b). STEREO EUVI 195 \AA~showing the hot ejecta and CME at a later time.}
\label{}
\end{figure}
\begin{figure}
\begin{center}
\vspace*{-2cm}
\hspace*{-0.4cm}
\hbox{
\includegraphics[width=0.9\textwidth,clip=]{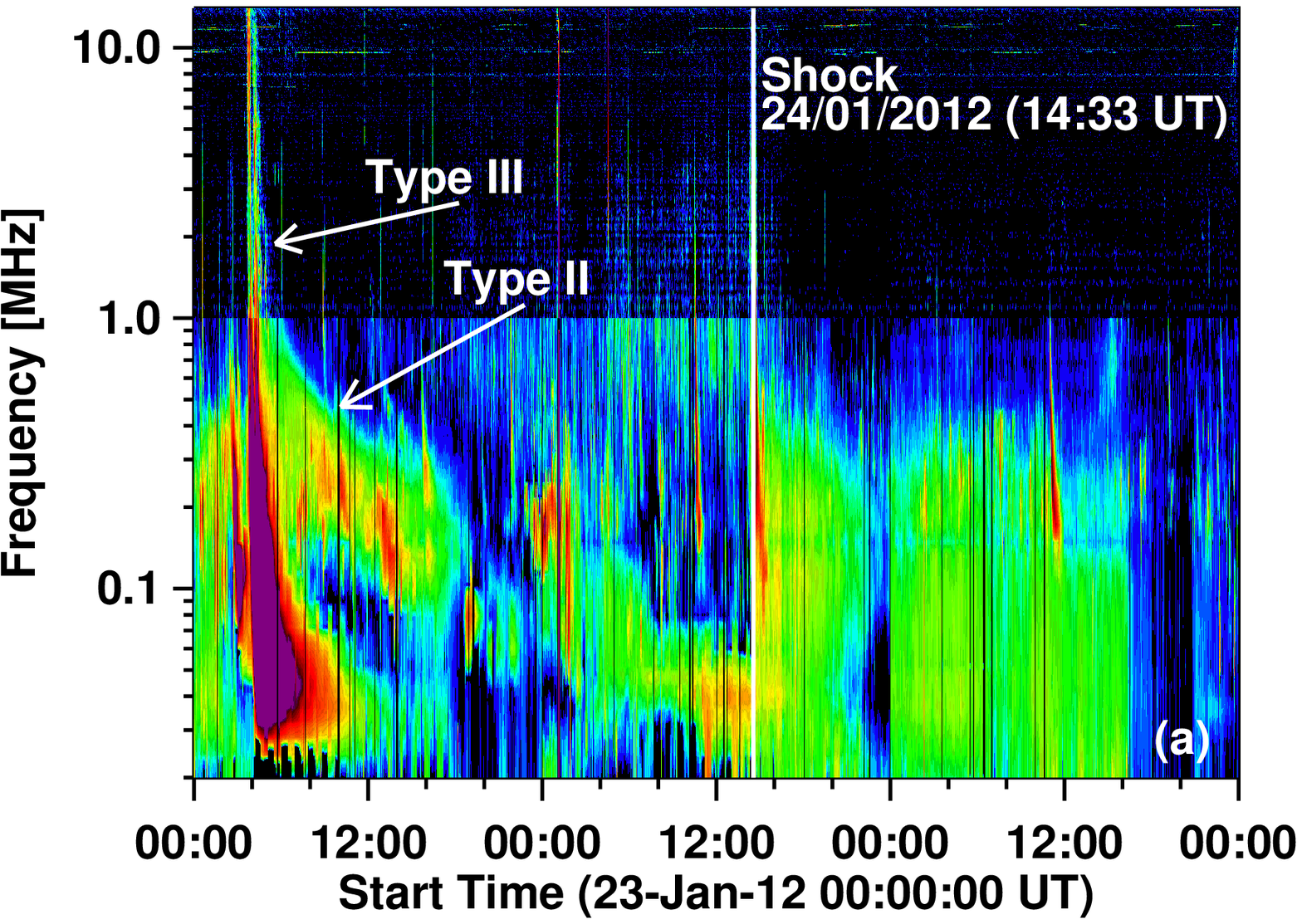}
}
\vspace*{-0.1cm}
\hspace*{-0.4cm}
\hbox{
\includegraphics[width=0.47\textwidth,clip=]{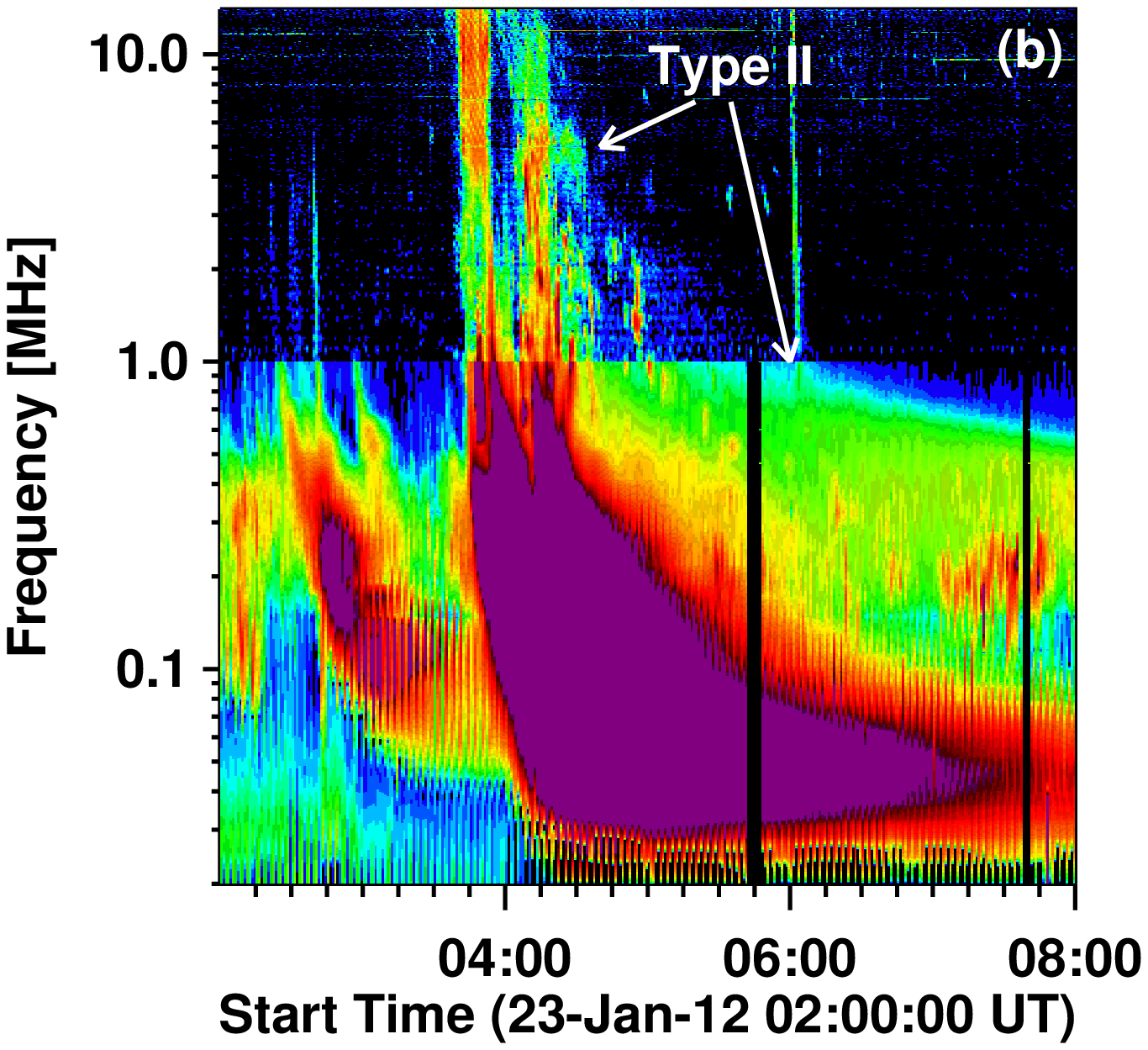}
\includegraphics[width=0.65\textwidth,clip=]{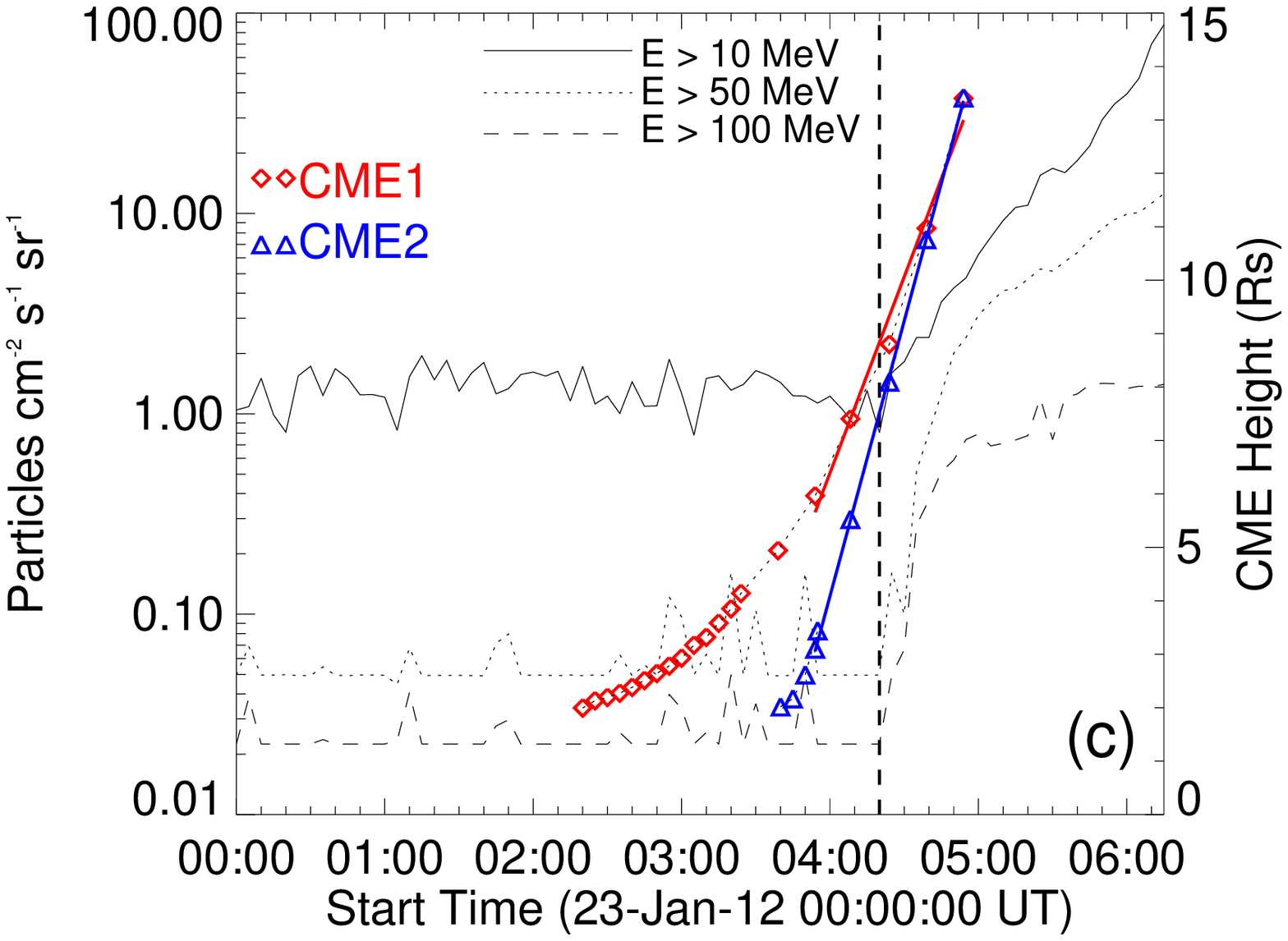}
}
\vspace*{0.5cm}
\hspace*{0.5cm}
\hbox{
\includegraphics[width=0.85\textwidth,clip=]{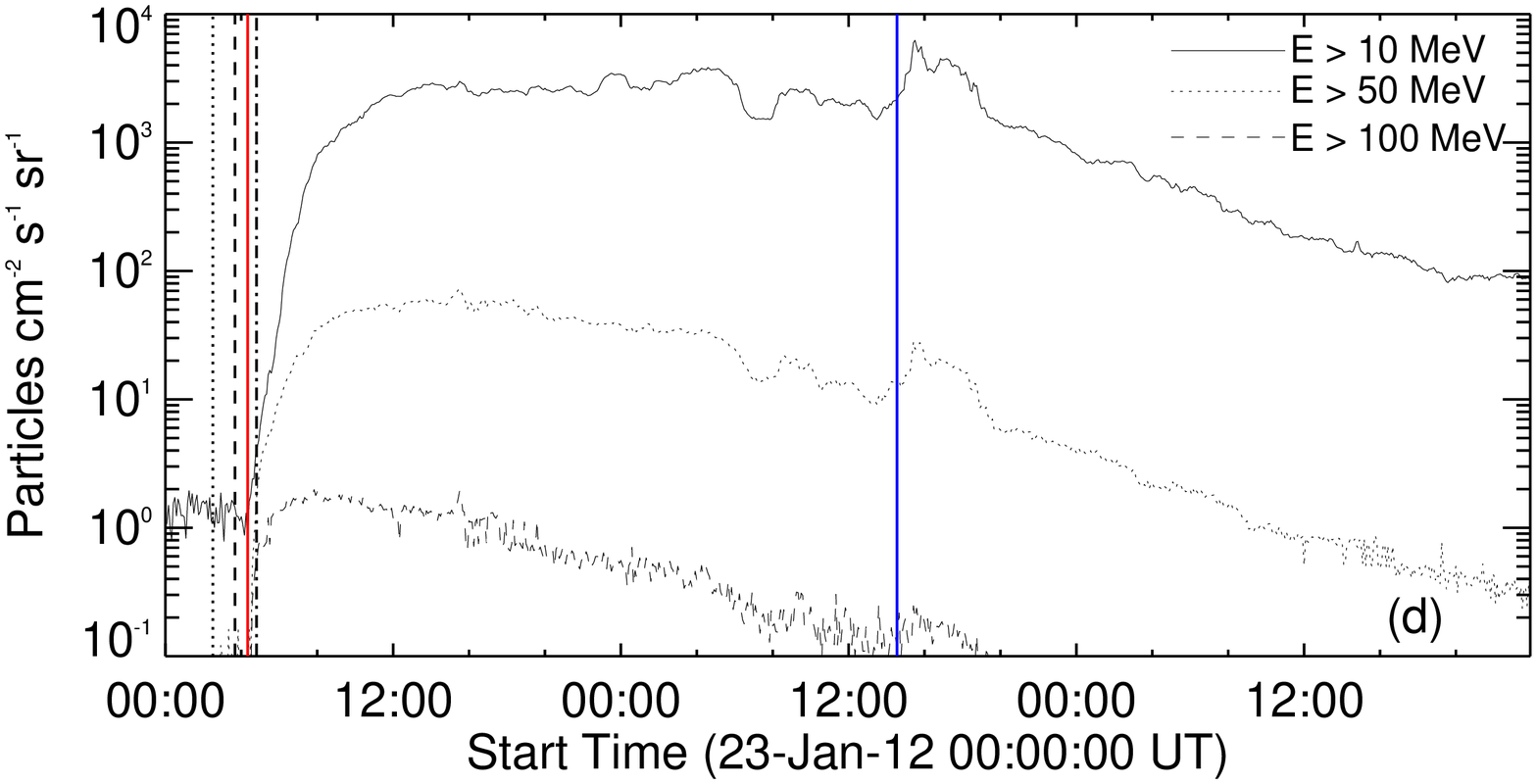}
}
\end{center}
\vspace*{-0.8cm} \caption{(a). Wind/WAVES dynamic radio spectrum showing interplanetary type II extending from the outer corona through the interplanetary space to the observing spacecraft. (b). same as (a) but displayed over a shorter interval (0-8 UT). (c). The SEP intensity profile in three energy channels ($>10$ MeV, $>50$ MeV, and $>100$ MeV) overplotted by the CME trajectories from 00:00 UT to $\sim$06:12 UT. (d). Same as (c) but for the three days extended time i.e., form January 23-25, 2012. The onset time of the CME1 and CME2 as it just appeared in the STEREO COR1 FOV, the onset of SEP event and the interaction end time of these CMEs has been presented by the vertical dotted, dashed, red and dashed dotted lines respectively.
}
\label{}
\end{figure}
\begin{figure}
\begin{center}
\hbox{
\includegraphics[width=0.5\textwidth,clip=]{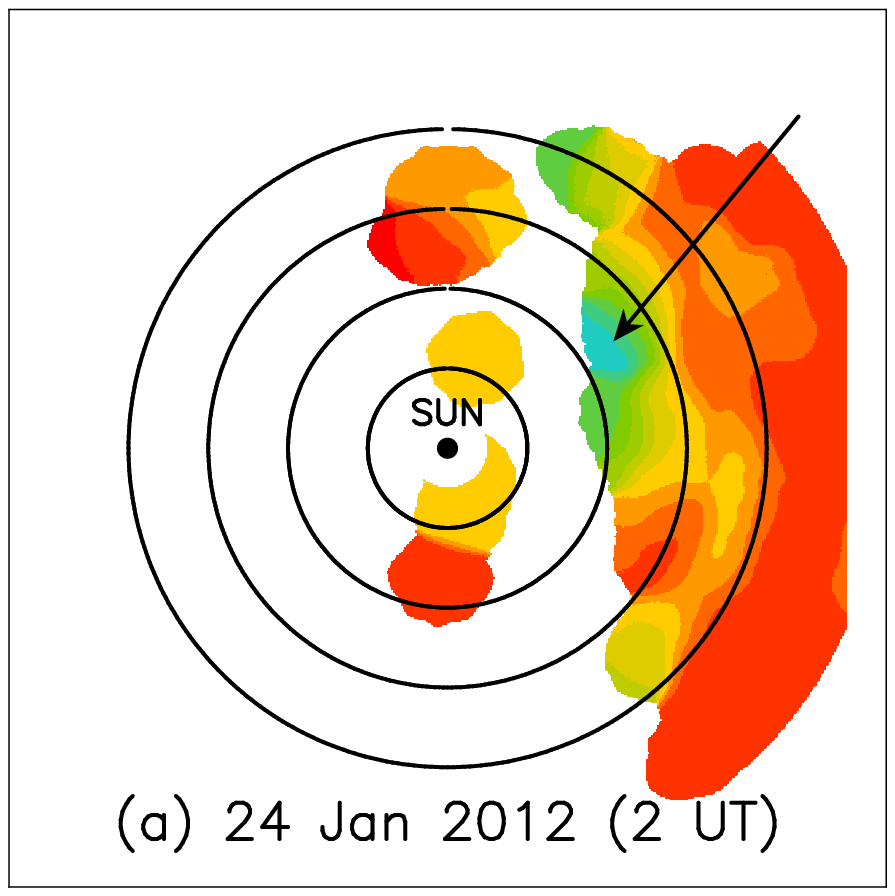}
\includegraphics[width=0.5\textwidth,clip=]{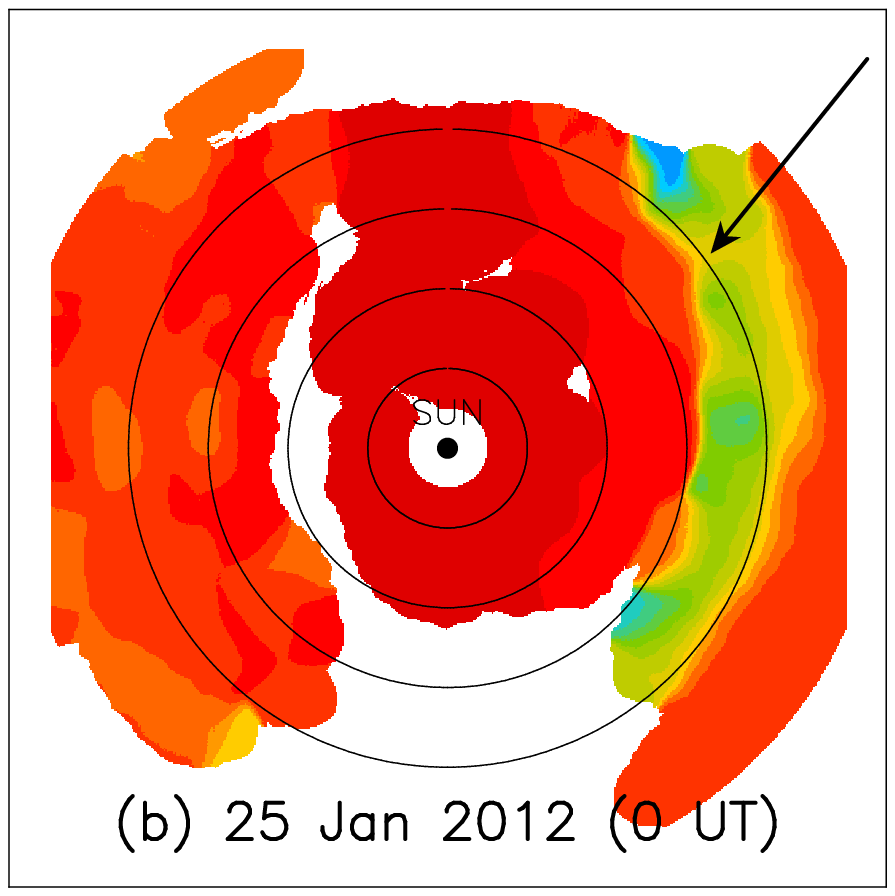}
}
\end{center}
\vspace*{-0.8cm} \caption{(a-b). Ooty scintillation images for January 24 and 25, 2012 measurements showing the 
propagation of CME into interplanetary space. In these 'PA - heliocentric distance' images, the north is at the top and east is to the left. The concentric circles are of radii 50, 100, 150 and 200 solar radius. The red color code indicates the background solar wind. In these images the observing time increases from right (west of the Sun) to left (east of the Sun). The arrows point to the compound CME at two different heights.
}
\label{}
\end{figure}
\begin{figure}
\begin{center}
\hbox{
\includegraphics[width=1\textwidth,clip=]{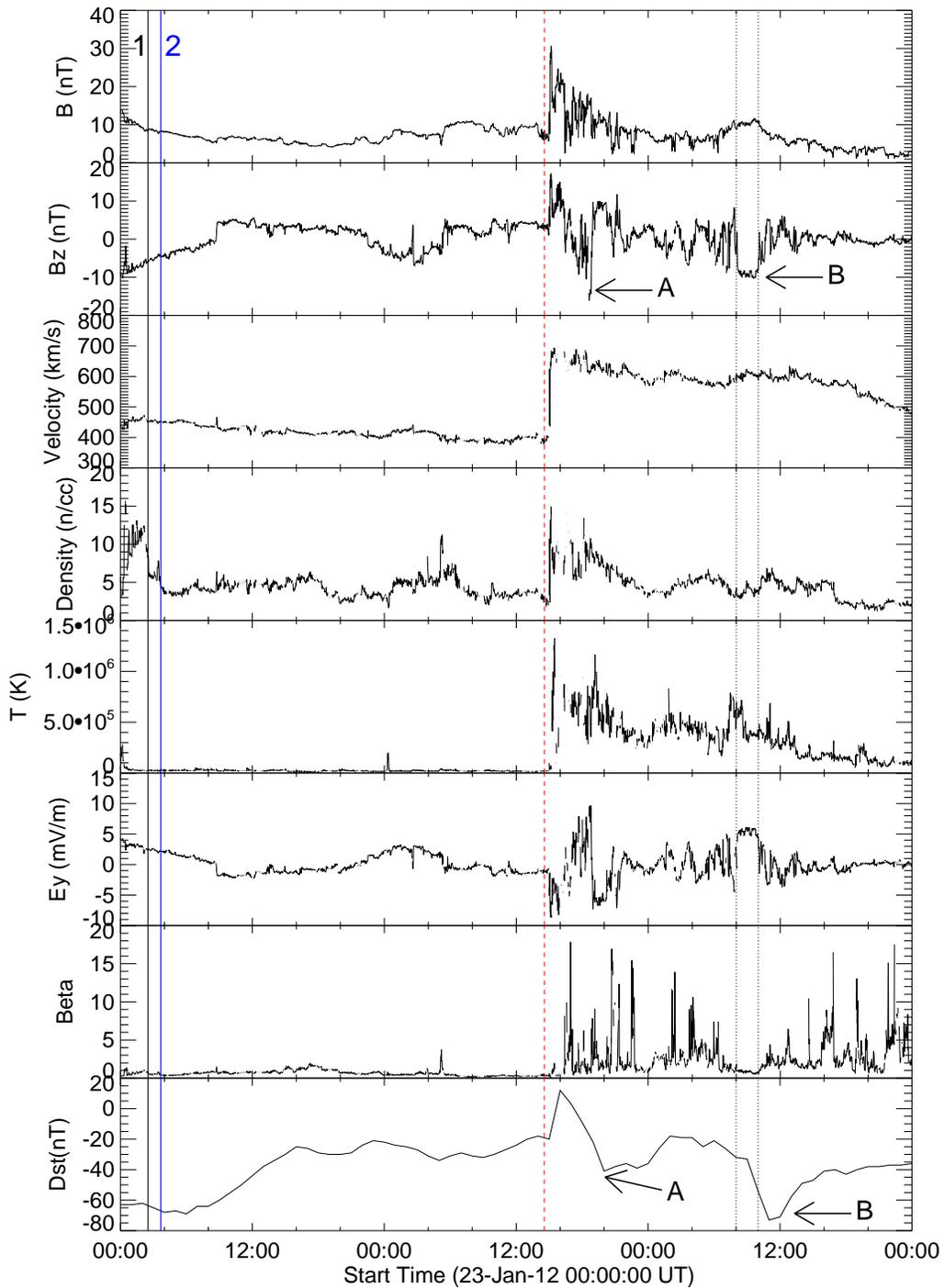}
}
\end{center}
\vspace*{-0.8cm} \caption{Plots of the average magnetic field (B) with its Z component, plasma flow speed (V), proton density (N), Proton Temperature (T), Electric field (Ey) (Data taken from OMNI data center) and Dst index. Lines 1 and 2 mark the onset of CME1 and CME2 in the STEREO-A COR1 FOV near the Sun. The vertical dashed line represents the arrival of the shock at Earth. Arrows A and B point to regions of negative Bz and the corresponding dips in the Dst index. The interval of negative Bz pointed by arrow A is due to the shock sheath and that pointed by arrow B is due to the ICME. The ICME boundaries are marked by the two vertical lines at 08:00 UT and 10:00UT on January 25, 2012.}
\label{}
\end{figure}
\end{document}